\documentclass[useAMS,usenatbib]{mn2e}
\usepackage{float}
\usepackage{graphicx}
\usepackage{amssymb}
\usepackage{amsmath}
\usepackage{datetime}
\usepackage[normalem]{ulem}
\usepackage{times}
\usepackage[export]{adjustbox}
\usepackage{soul}

\usepackage{color}
\usepackage{soul}
\definecolor{stan}{rgb}{0,0,1}

\definecolor{steve}{rgb}{0,1,0}

\newcommand{\araa}{Ann. Rev. Astron. Astrophys.}

\newcommand{\apj}{ApJ}
\newcommand{\apjs}{ApJS}
\newcommand{\apjl}{ApJ}
\newcommand{\mnras}{MNRAS}
\newcommand{\aap}{A\&A}

\newcommand{\apss}{Ap\&SS}

\newcommand{\jgr}{J. Geophys. Res.}

\newcommand{\solphys}{Sol. Phys.}

\newcommand{\ssr}{Sp.~Sci.~Rev.}
\newcommand{\azh}{Astron.~Zeitschrift}
\newcommand{\prl}{Phys.~Rev.~Lett.}
\newcommand{\grl}{Geophys.~Res.~Lett.}

\newcommand{\rA}{r_\mathrm{A}}

\newcommand{\RA}{R_\mathrm{A}}
\newcommand{\RK}{R_\mathrm{K}}
\newcommand{\hK}{h_\mathrm{K}}
\newcommand{\gK}{g_\mathrm{K}}
\newcommand{\vK}{v_\mathrm{K}}
\newcommand{\BK}{B_\mathrm{K}}
\newcommand{\DK}{D_\mathrm{K}}
\newcommand{\tauK}{\tau_\mathrm{K}}
\newcommand{\tauff}{\tau_\mathrm{ff}}
\newcommand{\sigdotK}{{\dot \sigma}_\mathrm{K}}
\newcommand{\sigdn}{\sigma_\mathrm{dn}}
\newcommand{\sigDn}{\sigma_\mathrm{Dn}}
\newcommand{\sigd}{\sigma_\mathrm{d}}
\newcommand{\sigD}{\sigma_\mathrm{D}}
\newcommand{\tsigd}{{\tilde \sigma}_\mathrm{d}}
\newcommand{\tsigD}{{\tilde \sigma}_\mathrm{D}}
\newcommand{\tsigDd}{{\tilde \sigma}_\mathrm{Dd}}

\newcommand{\Vesc}{V_\mathrm{esc}}

\newcommand{\vs}{v_\mathrm{s}}

\newcommand{\beq}{\begin{equation}}
\newcommand{\eeq}{\end{equation}}
\newcommand{\beqa}{\begin{eqnarray}}
\newcommand{\eeqa}{\end{eqnarray}}

\title[Leakage of centrifugal magnetospheres]{
Diffusion-plus-drift models for the mass leakage from centrifugal magnetospheres of magnetic hot-stars
}
\author[Owocki and  Cranmer]{Stanley P.\ Owocki$^1$
and Steven R.\ Cranmer$^2$
\\
$^1$Bartol Research Insitute, 
Department of Physics \& Astronomy, 
University of Delaware, Newark, DE 19716 USA
\\
$^2$Department of Astrophysical and Planetary Sciences, Laboratory for Atmospheric and Space Physics, University of Colorado,
\\ Boulder, CO 80309, USA 
}

\begin{document}

\include{aas_macros}

\date{Accepted ?.  Received ?; in original form ?}

\maketitle

\label{firstpage}

\begin{abstract}
In the subset of luminous, early-type stars with strong, large-scale magnetic fields and moderate to rapid rotation, material from the star's radiatively driven stellar wind outflow becomes trapped by closed magnetic loops, forming a centrifugally supported, co-rotating magnetosphere.
We present here a semi-analytic analysis of how this quasi-steady accumulation of wind mass can be balanced by losses associated with a combination of an outward, centrifugally driven {\em drift} in the region beyond the Kepler co-rotation radius, and an inward/outward {\em diffusion} near this radius.
We thereby derive scaling relations for the equilibrium spatial distribution of mass, and the associated emission measure for observational diagnostics like Balmer line emission.
We discuss the potential application of these relations for interpreting surveys of the emission line diagnostics for OB stars with centrifugally supported magnetospheres.
For a specific model of turbulent field-line-wandering rooted in surface motions associated with the iron opacity bump, we estimate values for the associated diffusion and drift coefficients.
\end{abstract}


\section{Introduction and Background}

Hot luminous, massive stars of spectral type O and B have dense, high-speed stellar winds, driven by line-scattering of the star's radiation
\citep[hereafter CAK]{Castor75}.
In the subset ($\sim$10\%) of massive stars with strong ($10^2 -10^4$\,G), globally ordered (often significantly dipolar) magnetic fields
\citep{Petit13}, the trapping of the wind outflow by closed magnetic loops leads to the formation of a circumstellar magnetosphere.
For stars with moderate to rapid rotation -- such that the Keplerian corotation radius $\RK$ lies within the Alfv\'{e}n radius $\RA$ that characterizes the maximum height of closed loops --, the rotational support leads to formation of a ``centrifugal magnetosphere'' (CM), wherein the trapped wind material accumulates into a relatively dense, stable and long-lived rigidly rotating disk \citep[][hereafter TO05]{Townsend05}.

A key outstanding question is how the mass feeding of such a CM is ultimately balanced by mass loss.
As analyzed in the appendices of TO05, as mass is added to a CM there comes a point where the magnetic tension becomes insufficient to confine the material against centrifugal breakout. This suggests a hierarchy of mass ejection events that limit the CM mass to an asymptotic value  (given by TO05 eqn. A11) that depends on field strength and stellar parameters, but is independent of mass feeding rate.
However, analysis of MOST observations of $\sigma$\,Ori\,E indicate that the absorption depth from magnetic clouds requires a magnetospheric density that is much less, by about a factor 50, than the values associated with such maximum mass from breakout \citep{To13}.
Moreover, space-based photometric monitoring of $\sigma$~Ori~E shows the continuum variation due to occultation by magnetic clouds is very steady and periodic, without any evidence for breakout events.
The overall inference is thus that the magnetosphere must be undergoing a more continuous mass leakage \citep{Carciofi13,Shultz16a,Shultz16b}.

\begin{figure}
\begin{center}
\includegraphics[scale=.65]{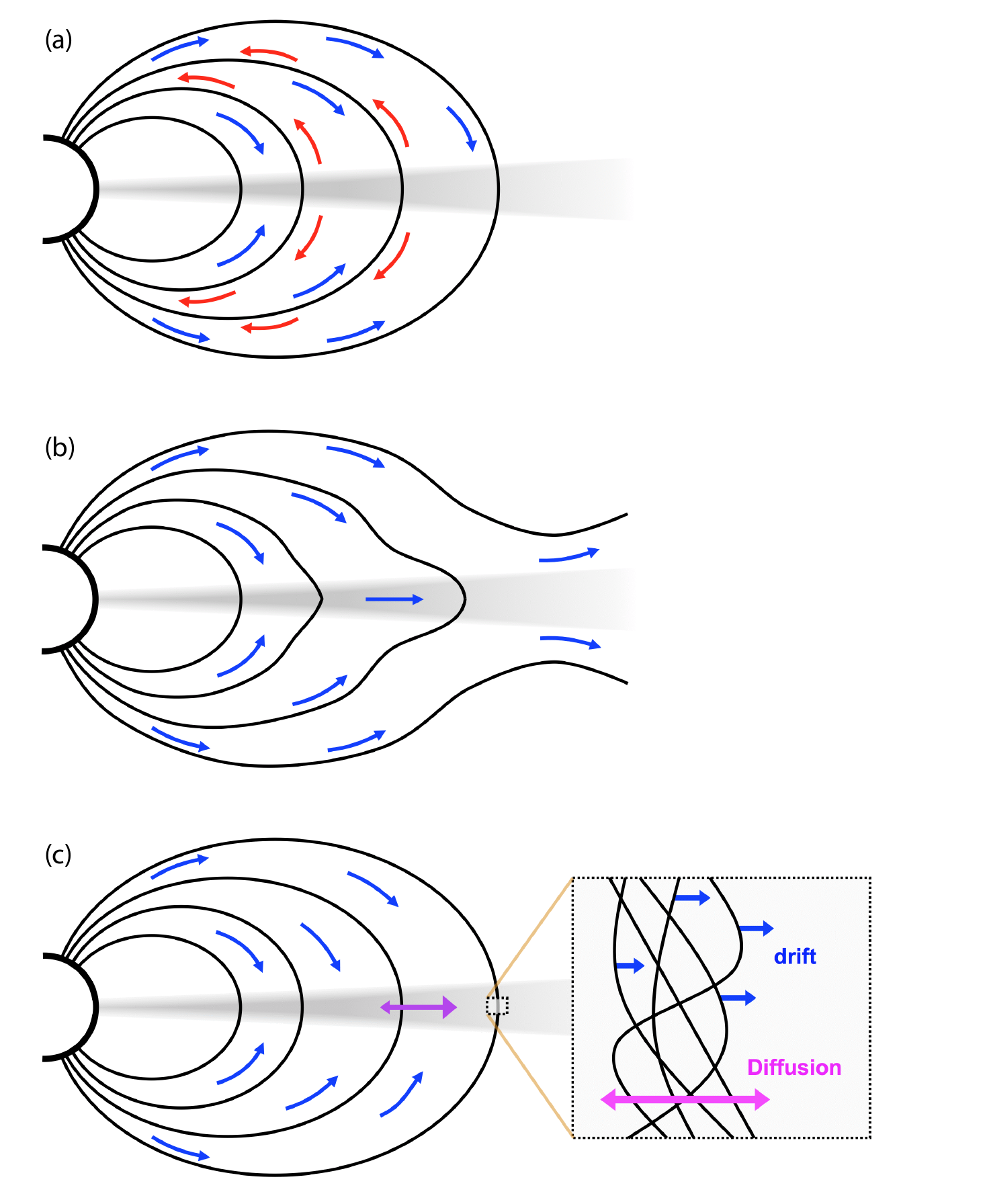}
\caption{Modes for emptying a stellar magnetosphere (gray shaded region) fed by a stellar wind mass upflow (blue arrows).
For case (a) with negligible rotation, wind upflows from opposite footpoints along a closed loop are stopped by collision near the loop top, forming a dense compression (grey region), with material then falling back to stellar surface in gravitational downflows (red arrows).
For case (b) with rotation, centrifugal support  against downfall causes material to build up to a high density, for which the total centrifugal force overcomes the magnetic tension, leading to a centrifugal breakout.
For case (c), non-ideal  terms associated with MHD turbulence along the field result in gradual mass leakage through a combination of inward/outward diffusion and outward drift of material.
Observations of centrifugally supported magnetospheres indicate a steady leakage akin to case (c), with densities too low for breakout mode of case (b).
The analysis in this paper quantifies the drift+diffusion processes of this case (c).
This graphic is patterned after figure 1 of \citet{HG84}.
}
\label{fig:hg84a}
\end{center}
\end{figure}

Figure \ref{fig:hg84a} here illustrates three distinct scenarios for magnetospheric mass loss, as identified in the insightful analysis by \citet[][see their figure 1]{HG84}.
The top panel shows a non-rotating or slowly rotating case, leading to a ``dynamical magnetosphere'' (DM) in which outflowing wind material trapped in closed magnetic fields undergoes shock compression near the loop top, and then cools and falls back to the star on a dynamical timescale; 
as detailed by both numerical simulations \citep{udDoula02} and analytic analyses \citep{Owocki16}, the rapid leakage from such dynamical downfall limits the buildup of mass density in such a DM.
The middle panel illustrates the case of centrifugally driven breakout (TO05), which as noted above requires a density that significantly exceeds that inferred from observations
\citep{To13}.

The lowermost panel of figure \ref{fig:hg84a} illustrates the case explored in this paper;
as highlighted by the callout box,  the wind feeding of mass into the closed loops is balanced by a combination of  
{\em diffusion} inward and outward from the Kepler co-rotation radius $\RK$, plus a centrifugally driven, outward {\em drift} of material in the regions above $\RK$.
This diffusion+drift formalism is developed in \S 2, while \S 3 derives the implied scalings for CM density and emission measure.
\S 4 discusses the evaluation of diffusion and drift coefficients for a model based on magnetohydrodynamic (MHD) turbulence.
We conclude in \S 5 with a discussion of the applicability of our results for interpreting observational diagnostics, together with an outlook for future work.
 
\section{General mass balance}

\subsection{Wind feeding rate of disk}

For analytic tractability, we focus our analysis here on the axisymmetric case of a rotation-aligned dipole, for which the mass accumulation surface is an equatorial disk ranging roughly from the Kepler corotation radius $\RK$, where the centrifugal forces from rigid-body corotation balance gravity with the speed of a Keplerian orbit, to the Alfv\'{e}n radius $\RA$, where the closed loops of the assumed stellar dipolar fiield are forced open by the stellar wind outflow.
At a local loop-apex radius $R$, the wind feeding rate of the disk is
\citep{OU04}
\beq
{\dot \sigma} (R) = \frac{{\dot M}}{2 \pi R_\ast^2} \, \frac{2 \mu_\ast (R)}{\sqrt{1+3 \mu_\ast^2}} \frac{B(R)}{B_\ast} 
\, ,
\label{eq:sigdef}
\eeq
where ${\dot M}$ is the CAK mass loss rate for a non-magnetized star, $B(R)$ and $B_\ast$ are respectively the field strengths at the disk radius $R$ and at the loop footpoint at stellar radius $R_\ast$,
and 
the term involving the co-latitude $\mu_\ast (R) = \sqrt{1-R_\ast/R}$ represents the local surface tilt angle of the field
 \citep[cf.\ eqn.\ 6 of][]{Owocki16}.
 For simplicity, we ignore here the effect of rotation on the  shape of the underlying star.
 
\begin{figure}
\begin{center}
\includegraphics[scale=.65]{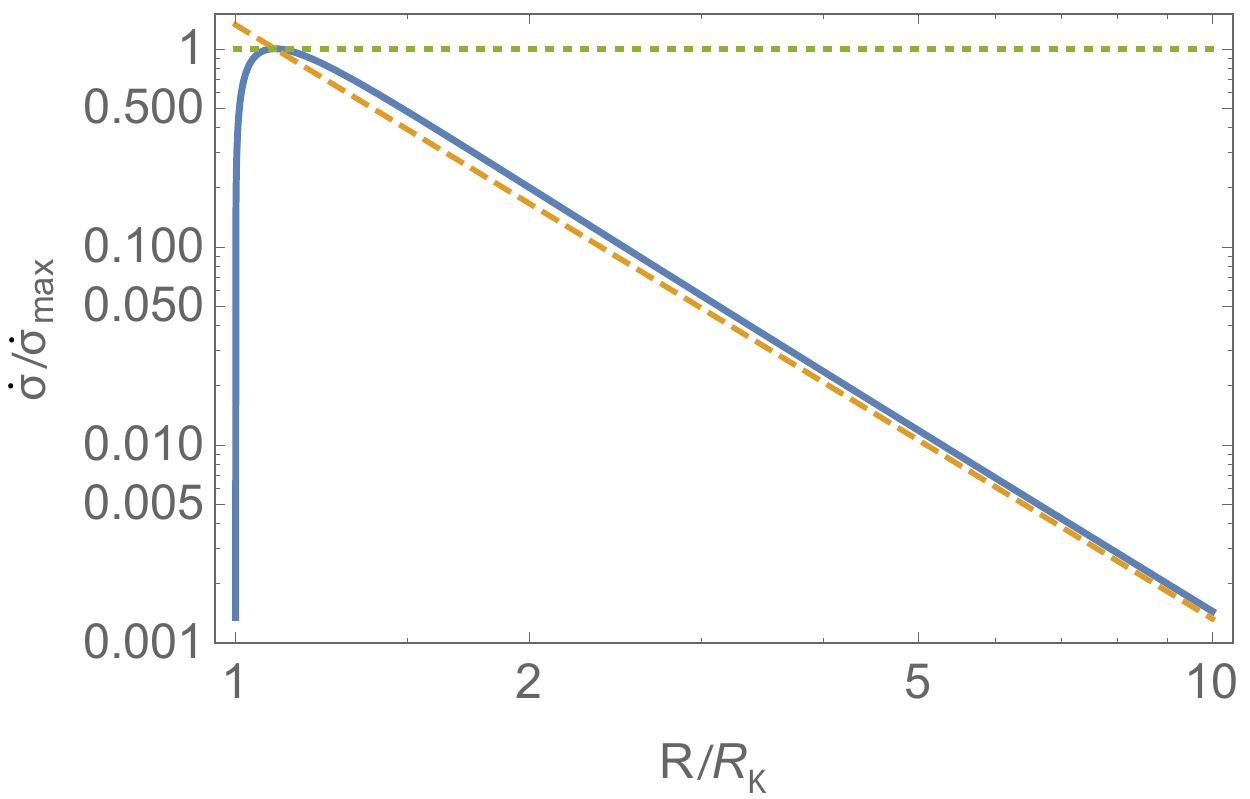}
\caption{
Radial variation of disk feeding rate from the stellar wind, normalized to its peak value $\dot{\sigma}_{\rm max}$, which occurs at $R_{\rm peak} = 1.097 R_\ast$.
The red diagonal dashed line compares the simple form $(R_{\rm peak}/R)^3$.
The green horizontal dotted line simply marks the value unity.
}
\label{fig:sigdotr}
\end{center}
\end{figure}

Using the dipole form for equatorial field strength,
\beq
\frac{B(R)}{B_\ast} = \left ( \frac{R_\ast}{R} \right )^3 \, \frac{1}{\sqrt{1+ 3 \mu_\ast^2}}
\, ,
\eeq
the disk feeding rate can be written as an explicit function of radius,
\beqa
{\dot \sigma} (R) &=& \frac{{\dot M}}{2 \pi R_\ast^2} \left ( \frac{R_\ast}{R} \right )^3 \, \frac{2 \sqrt{1-R_\ast/R}}{4-3R_\ast/R} 
\nonumber
\\
&\approx&  \frac{{\dot M}}{4 \pi R_\ast^2} \left ( \frac{R_\ast}{R} \right )^3  ~~ ; ~~ R \gg R_\ast
\, ,
\label{eq:sigr}
\eeqa
where the latter expression provides a suitable approximation for most CM cases of interest, i.e. with $\RK > 1.5 R_\ast$.
Figure \ref{fig:sigdotr} plots the radial variation of $\dot{\sigma}$, normalized to its peak value at $R_{\rm peak} =1.097 R_\ast$.

From TO05 eqns.\ (18) and (28), the Gaussian scale height of the disk can be written as
\beqa
h(R) &=& \RK \frac{\vs}{\vK} \, \frac{\sqrt{2}}{\sqrt{3-2(\RK/R)^3}}
\nonumber
\\
 &=& \frac{\vs}{\Omega} \, \frac{\sqrt{2}}{\sqrt{3-2(\RK/R)^3}}
\nonumber
\\
 &\equiv&  \frac{\hK}{\sqrt{3-2(\RK/R)^3}}
\, ,
\label{eq:hr}
\eeqa
where $\Omega$ is the angular rotation frequency of the star, and $\vK=\Omega \RK$ is the common orbital/rotation speed at the Kepler radius $R_{\rm K} \equiv \left ( GM/\Omega^2 \right )^{1/3}$, with $M$ the stellar mass and $G$ the gravitation constant. In addition,
$\vs$ is the isothermal sound speed, and $\hK \equiv \sqrt{2} \, \vs/\Omega$ is the disk thickness at the Kepler radius.
Since $ \vs \ll \vK$, the disk is generally very thin, $h \ll r $, with the gas volume density $\rho$ having a narrow Gaussian stratification about a central disk-center value $\rho(R,0)$, giving a disk surface density $\sigma(R) \approx \rho(R,0) h(R)$.

In TO05, the radial variation of surface density was simply taken to be proportional to the disk feeding rate, effectively assuming accumulation for a fixed time, associated with a presumed emptying of the entire magnetosphere due to centrifugal breakout.
However, as noted in the introduction, there is now evidence that the overall density and mass of CM's is well below what is needed for such a large-scale emptying by centrifugal breakout.
The overall implication is thus that the magnetosphere must be undergoing a more continuous mass leakage, due to diffusion and/or drift of material across closed field lines.
The following sections derive expressions for the equilibrium surface density that arises from a balance of the wind feeding against magnetospheric leakage from a combination of diffusion and drift.

\subsection{Balance between mass feeding and divergence of mass flux}

In steady-state, mass conservation requires that the divergence of the radial mass flux $F$ in the disk is balanced by the local disk feeding rate $\dot{\sigma} (R)$.
Adopting a cylindrical coordinate system with axial radius $r \equiv R/\RK$ scaled by the Kepler radius $\RK$,  this mass conservation condition can be written as
\beq
\frac{1}{r} \frac{d(r F) }{d r}=  \sigdotK \RK  \, r^{-3}
\, ,
\label{eq:diffeq}
\eeq
where $\sigdotK \equiv {\dot \sigma} (\RK)$ is the disk feeding rate at the Kepler radius, and the inverse cubic scaling with radius follows from eqn.\ (\ref{eq:sigr}).
This can be readily integrated to yield
\beq
F(r) =\frac{\sigdotK \RK}{{\it r}} \,  \left ( C_1-\frac{1}{r} \right )
\, ,
\label{eq:Fr}
\eeq
where $C_1$ is an integration constant.
For example, for a boundary condition that the mass flux vanishes at the Kepler radius, i.e., $F(1)\equiv 0$,  we require $C_1 =1$.

This disk mass flux $F$ could arise from a {\em diffusion} associated\footnote{Here we adopt the notational convention that subscripts
with capital ``D" represent diffusion, while those with lower case ``d" refer to drift.} with a radial gradient in surface density $\sigma$,
\beq
F_D = - \frac{D}{\RK} \frac{d\sigma}{dr} =  - \frac{\DK}{\RK}  r^a \sigma' (r)
\, ,
\label{eq:FD}
\eeq
where the latter equality invokes the usual shorthand $ \sigma' (r) \equiv d\sigma/dr$, and introduces an assumption that the diffusion coefficient $D$ varies in radius as a power-law of index $a$ away from its value $\DK$ at the Kepler radius.
For example, for diffusion that scales just with the inverse of the magnetic field strength, $D(r) = \DK \BK /B(r) \sim r^3$, so that $a=3$.

Beyond the Kepler radius, there arises an effective net outward centrifugal-gravitational acceleration
\beq
g(r) = \gK \left ( r - \frac{1}{r^2} \right )
\, ,
\label{eq:gnet}
\eeq
where the gravitational acceleration at the Kepler radius, $\gK \equiv GM/\RK^2 = \Omega^2 \RK$, with $\Omega$ the star's angular rotation frequency.
For acceleration limited by an effective `collision time' $\tau$, this leads to a net outward {\em drift} flux,
\beq
F_{\mathrm d} = \sigma \, g \, \tau = \sigma \gK \tauK r^b \, \left ( r - \frac{1}{r^2} \right )
\, ,
\label{eq:Fd}
\eeq
where the latter equality now introduces an assumption that this collision time $\tau$ varies in radius as a power-law of index $b$ away from its vaue $\tauK$ at the Kepler radius.
Again, if $\tau \sim 1/B(r) \sim  r^3$, then $b=3$.
Note that this drift flux vanishes at the Kepler radius, $F_d (1) = 0$.

Applying eqns.\ (\ref{eq:FD}) and (\ref{eq:Fd}) in eqn.\ (\ref{eq:Fr}), we obtain a general first-order differential equation for the surface density $\sigma$ arising from diffusion+drift transport of mass that is fed by the stellar wind at a rate ${\dot \sigma}$,
\beqa
 - \frac{\DK}{\RK}  r^{a+1} \sigma' (r)  &+&  
 \gK \tauK r^{b+1}  \left ( r - \frac{1}{r^2} \right ) \sigma(r) =
\nonumber
\\
&&~~~~~~~~~~~~~~~
  \sigdotK \RK \,  \left ( C_1-\frac{1}{r} \right )
\, .
\label{eq:Ddode}
\eeqa

\subsection{Diffusion with no drift}

Before considering solutions for the general case with both diffusion and drift, let us build insight by examining in turn models with only one of the effects, beginning first with pure diffusion. Setting $\tauK=0$ in eqn.\ (\ref{eq:Ddode}), we see that a characteristic normalization scale for the surface density in this diffusion case can be defined as
\beq
\sigDn \equiv \frac{\sigdotK \RK^2}{\DK}
\, .
\label{eq:Dn}
\eeq

We can readily integrate the pure-diffusion form of (\ref{eq:Ddode}) (with $\tauK=0$).
Since material accumulating near the Kepler radius can generally be expected to diffuse both outward and inward to lower density regions above and below, let us eliminate the integration constants by applying  boundary conditions of vanishing density at both the Kepler\footnote{As shown in TO05 eqn.\ (19), the innermost radius for centrifugal confinement can extend down to $r_i = \sqrt[3]{2/3} = 0.87$, but for simplicity, we ignore this small difference and assume an inner boundary condition that $\sigma(r_i=1)=0$.} and Alfv\'{e}n\footnote{The Alfv\'{e}n radius roughly marks the outermost radius for closed field lines, above which mass loss occurs by wind outflow that is much faster than the flux from diffusion or drift, with a density that is thus much lower than in the magnetically trapped disk; this motivates the zero density outer boundary condition used here.} radii, i.e. $\sigma(1)=\sigma(\rA) \equiv 0$. 
The normalized solution  takes the analytic form,
\beqa
\tsigD (r,a,\rA) &\equiv& 
\frac{\sigD (r,a,\rA) }
{\sigDn}
\nonumber
\\
&=&
\frac{\rA-r + \rA^{1+a}(r-1) -r^{1+a}(\rA-1)}
{r^{1+a} (1+a) \rA (\rA^a -1)} 
\, .
\nonumber
\\
\label{eq:sigbidiffr}
\eeqa
This peaks at a radius above the Kepler radius, given by
\beq
r_{\rm peak} = 
\frac
{(1+a) \rA (\rA^a - 1)}
{a (\rA^{1+a} -1) }
\, .
\label{eq;rpeak}
\eeq
with peak density ${\tilde \sigma}_D (r_{\rm peak},a,\rA) $.

\begin{figure}
\begin{center}
\includegraphics[scale=.65]{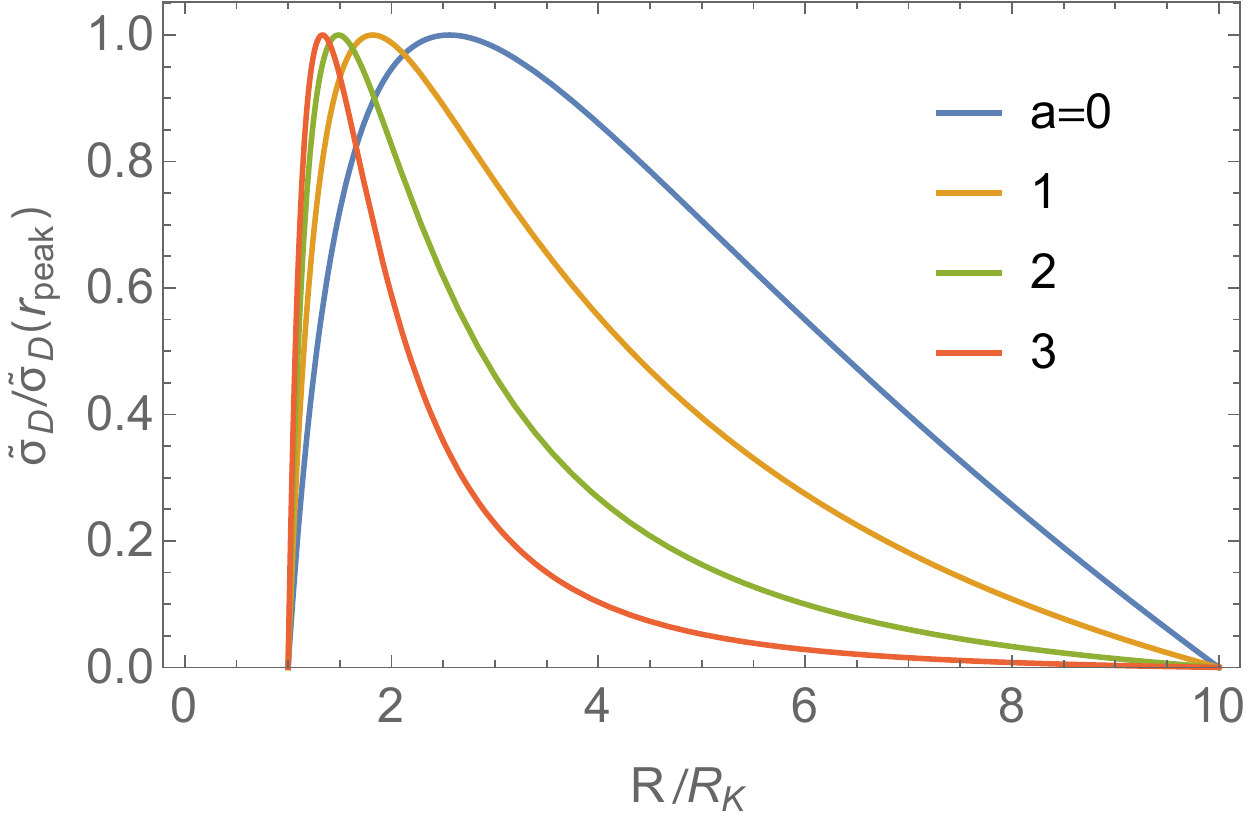}
\caption{
For the pure diffusion solution (\ref{eq:sigbidiffr}), radial variation of surface density, normalized by its peak value,
for Alfv\'{e}n radius fixed to $\rA= 10$ Kepler radii, but with various diffusion exponents $ a=0$, 1, 2, and 3 (with peaks ordered from right to left).
}
\label{fig:sigbidiffr}
\end{center}
\end{figure}

For a fixed Alfv\'{e}n radius $\rA=10$, figure \ref{fig:sigbidiffr} plots the radial variation of surface density, normalized by this peak value, for
various values of the diffusion exponent\footnote{For exponents $a=0$ and $a=-1$ the expression (\ref{eq:sigbidiffr}) diverges and so should be replaced with alternative forms involving logarithms. In practice, however, one can still obtain the functional variation by evaluating (\ref{eq:sigbidiffr}) using exponents $a$ that are just slightly displaced from these critical values.} $ a=3$, 2, 1, and 0.

\subsection{Drift with no diffusion}

For the case with drift but no diffusion, we can simply set $\DK=0$ in eqn.\ (\ref{eq:Ddode}) and solve directly for the density $\sigma$.
To keep the density finite at the Kepler radius, we must choose the integration constant $C_1=1$.
The overall normalization of the density now takes the form,
\beq
\sigdn \equiv \frac{\sigdotK \RK}{\gK \tauK} =  \frac{\sigdotK}{\Omega^2 \tauK} 
\, .
\label{eq:sigdn}
\eeq
The solution for the scaled density is then,
\beq
\tsigd (r,b,\rA) 
\equiv \frac{\sigd (r,b,\rA)}{\sigdn}
=  \frac{1}{r^b(r^2+r+1)}
\, .
\label{eq:sigtdriftr}
\eeq
At large radii $r \gg 1$ well above the Kepler radius, this just approaches a simple power-law decline, $\tsigd \sim 1/r^{b+2}$.

\subsection{Drift + Diffusion model}

Let us now explore general models with {\em both} diffusion and drift. 
To characterize the overall relative strength of drift to diffusion, let us define 
\beq
 \gamma \equiv
 \frac{\sigDn}{\sigdn} =
 \frac{\gK \tauK \RK}{\DK} =
 \frac{ \Omega^2 \RK^2 \tauK}{\DK } =
 (\Omega \tauK) ~ (\Omega \tau_{\mathrm D,K})
\, ,
\label{eq:gamdef}
\eeq
where 
$\tau_{\mathrm D,K} \equiv \RK^2/\DK$ is a characteristic diffusion time at the Kepler radius.
Scaling the density again by the diffusion normalization $\sigDn \equiv \sigdotK \RK^2/\DK$,
the general diffusion+drift eqn.\ (\ref{eq:Ddode}) takes the scaled form,
\beq
- r^{1+a} 
{\tilde \sigma '}_{\mathrm{Dd}} (r)
 + 
\gamma \, r^{1+b} \left ( r - \frac{1}{r^2} \right ) \tsigDd (r) = C_1 - 1/r
\, .
\label{eq:ddeqn}
\eeq

 With boundary conditions again that this density vanishes both at the Kepler and Alfv\'{e}n radii,
i.e.\  $\tsigDd(1)=\tsigDd(\rA) = 0$, eqn.\ (\ref{eq:ddeqn}) can be further integrated {\em numerically} to give $\tsigDd (r,\gamma,a,b,\rA)$.

But for the special case $b=a+1$, integration of (\ref{eq:ddeqn}) takes the {\em analytic} form,
\beqa
&& \tsigDd (r, \gamma, a,a+1) 
=
 e^{\frac{\gamma  r^3}{3}}  \left\{ {C_1}{r^{-\gamma }}   + \frac{r^{-a-1}}{3 } \times  \right. ~~~~~~
\nonumber
\\
&& ~
\left.
 \left[ C_2 \, r \,  E_{\frac{1}{3} (a-\gamma +3)}\left(\frac{r^3 \gamma
   }{3}\right)-E_{\frac{1}{3} (a-\gamma +4)}\left(\frac{r^3 \gamma }{3}\right)\right]  \right\}
\, ,
\nonumber
\\
\label{eq:sigddc1c2}
\eeqa
where $E_n (x)$ is the exponential integral. 
The boundary conditions $\tsigDd(1)=\tsigDd(\rA) = 0$ can then be used to eliminate the integration constants $C_1$ and $C_2$.
The resulting expression is straightforward to evaluate, but lengthy in form, so will not be explicitly written out here.

\begin{figure}
\begin{center}
\includegraphics[scale=.65]{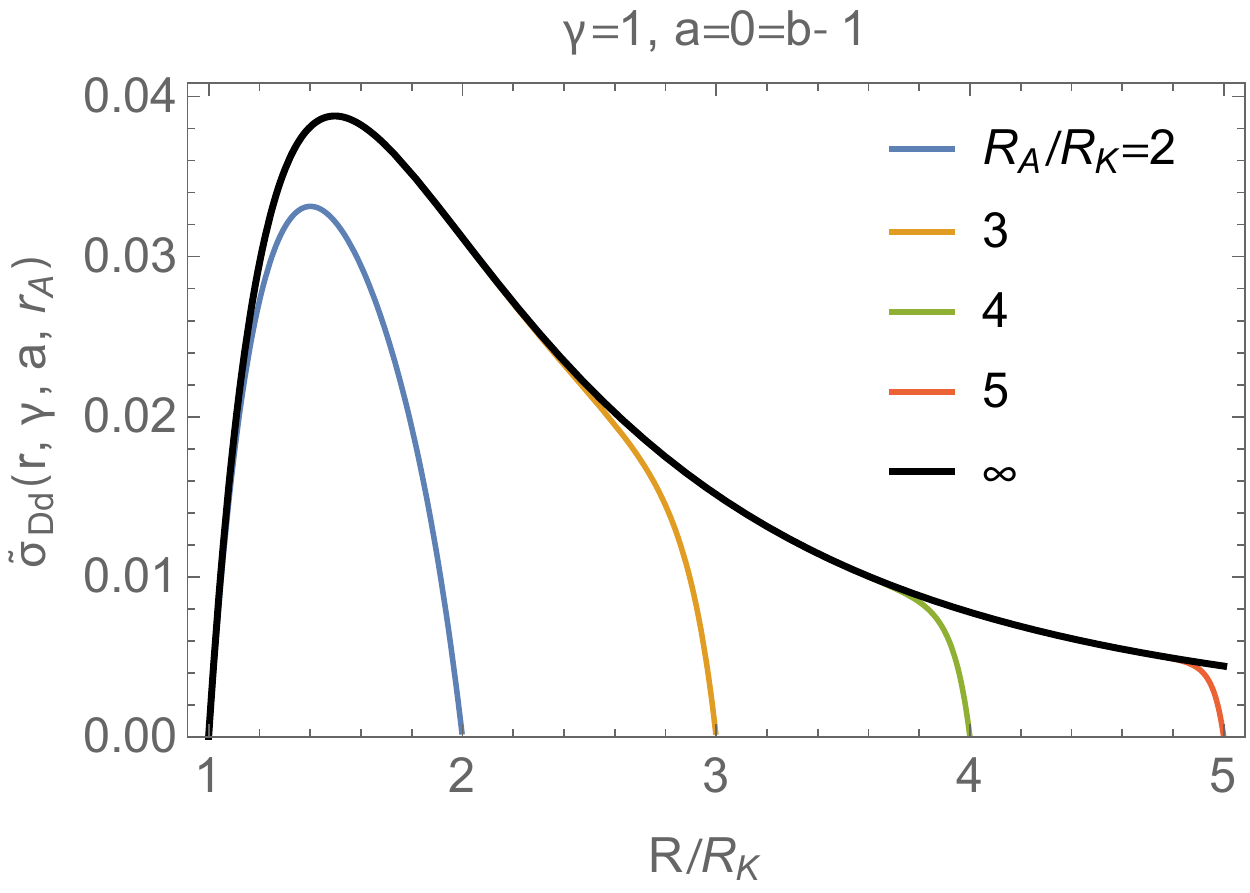}
\caption{Radial variation of scaled surface density $\tsigDd$ for diffusion+draft analytic solution with $\gamma=1$, $a=b-1=0$, and for labeled values of the ratio of the Alfv\'{e}n to Kepler radius.
}
\label{fig:dd-soln-g1a0ro}
\end{center}
\end{figure}

\begin{figure*}
\begin{center}
\includegraphics[scale=.65]{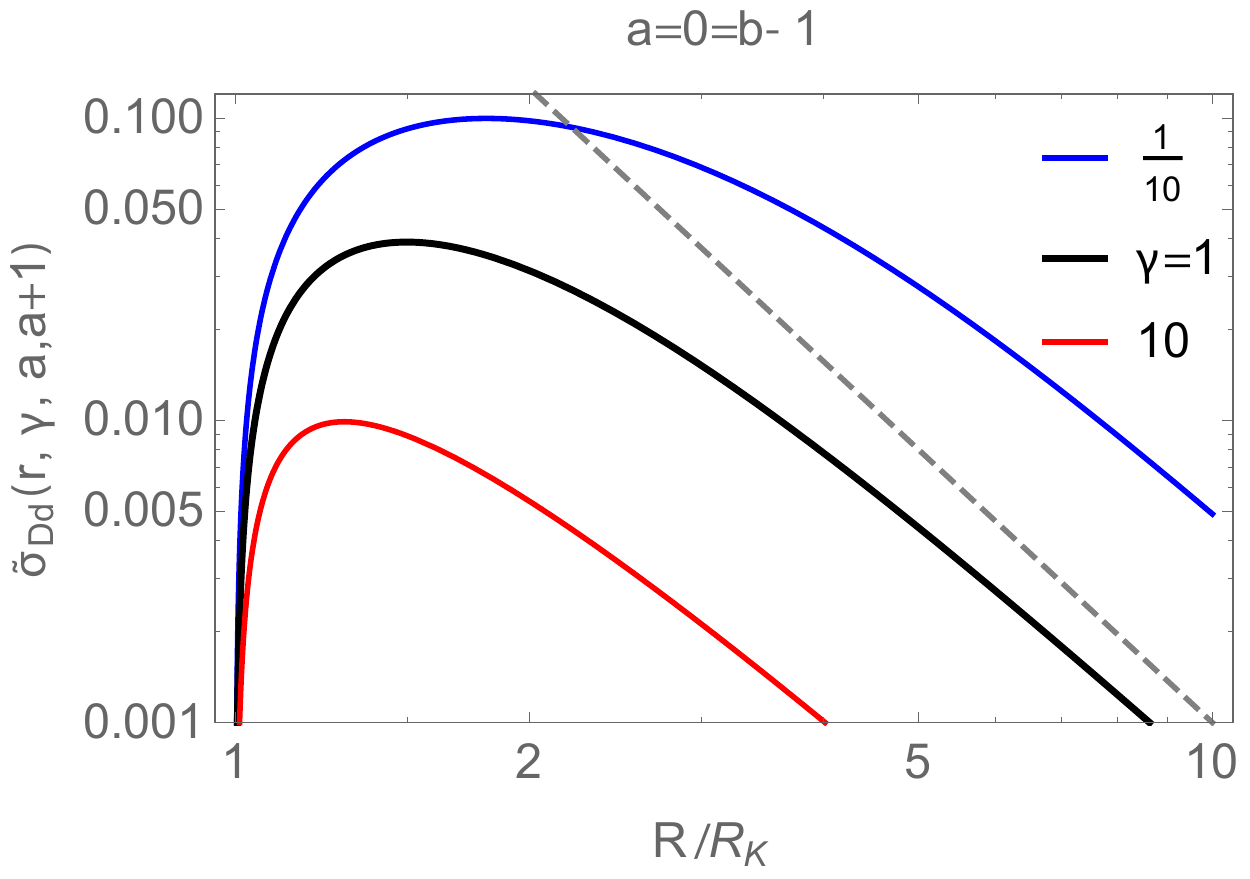}
~~
\includegraphics[scale=.65]{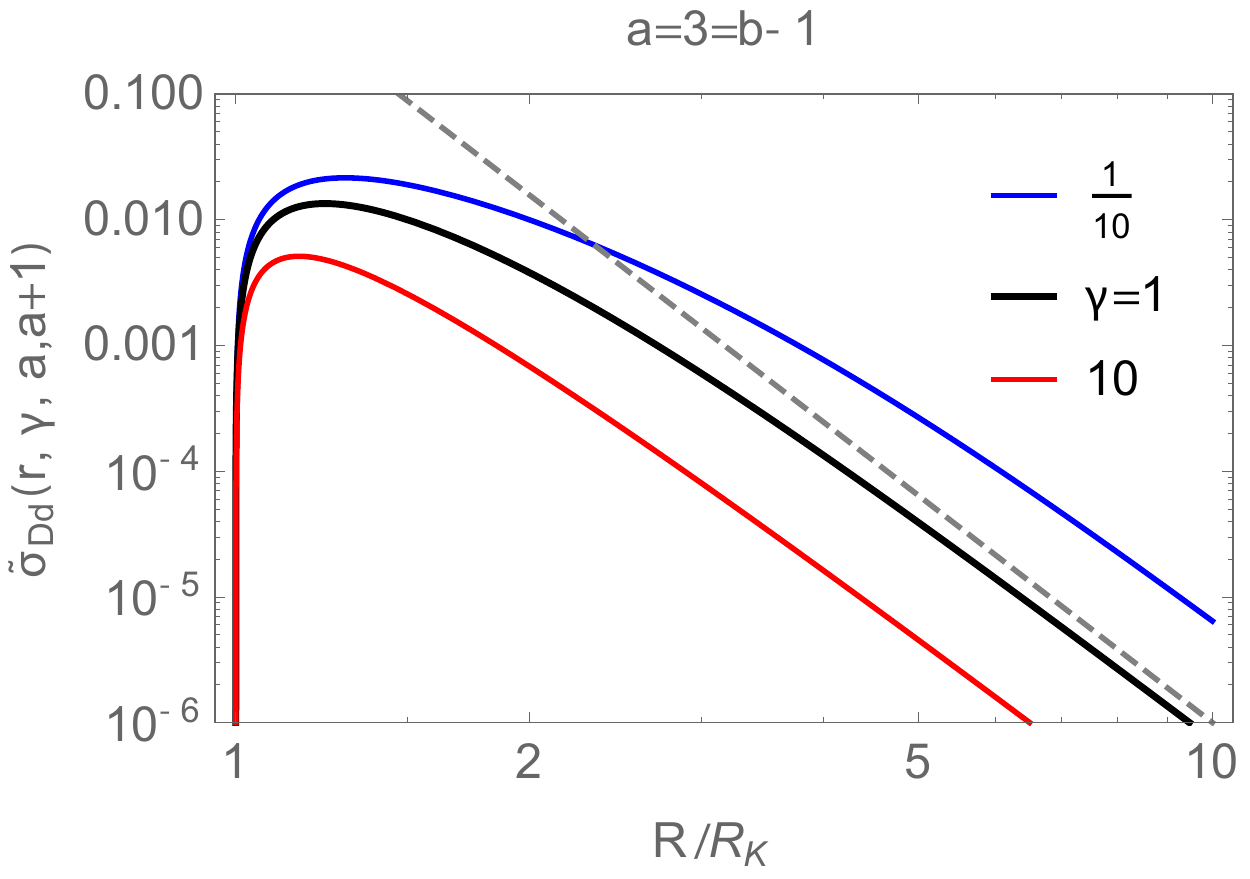}
\caption{Radial variation of normalized surface density $\tsigDd$ for analytic solutions (\ref{eq:sigddr2}) with diffusion and drift, for cases with $a=b-1=0$ (left) and $a=b-1=3$,  and for $\gamma$ values given by labels in upper right panel. 
The dashed lines show the power-law variation expected for drift at large radii, $\tsigDd \sim 1/r^{b+2} \sim 1/r^{a+3}$.}
\label{fig:dd-solns}
\end{center}
\end{figure*}

For a representative sample case ($\gamma=1$, $a=b-1=0$), figure \ref{fig:dd-soln-g1a0ro} plots the radial variation of scaled surface density for the labeled values of the  outer (Alfv\'{e}n) radius.
Except for the lowest case $\rA=2$, all the cases with a finite Alfv\'{e}n radius have a sharp cutoff near $\rA$, but otherwise fixed functional form, set by the thick black curve, representing the case with arbitrarily large Alfv\'{e}n radius.

Indeed, in this limit of $\rA \rightarrow \infty$, the fixed function for general $\gamma$ and $a=b-1$ 
takes the fully explicit analytic form,
\beqa
&&\tsigDd (r, \gamma, a,a+1) 
=
 \frac{1}{3} e^{\frac{\gamma  r^3}{3}}  r^{-a}
 \times 
\nonumber
\\
&&~~~~~~~~~~~~~~~~~ 
\left[\frac{E_{\frac{1}{3} (a-\gamma
   +4)}\left(\frac{\gamma }{3}\right) E_{\frac{1}{3} (a-\gamma +3)}\left(\frac{r^3 \gamma
   }{3}\right)}{E_{\frac{1}{3} (a-\gamma +3)}\left(\frac{\gamma
   }{3}\right)}  
   \right.
\nonumber
\\
&& ~~~~~~~~~~~~~~~~~~~~~~~~~
\left.
- ~ \frac{E_{\frac{1}{3} (a-\gamma +4)}\left(\frac{r^3 \gamma
   }{3}\right)}{r}\right]
\, .
\label{eq:sigddr2}
\eeqa

Figures \ref{fig:dd-solns} plot the radial variation of this normalized surface density for various cases.
The heavy black curve represents results for the $\gamma=1$ case of equal drift and diffusion strengths, 
while the colored curves show results for $\gamma=10$ (red) and $\gamma=1/10$ (blue).
The left panels show the case of constant coefficients ($a=0$), while the right panels show the $a=3$ case that would arise from $ D \sim \tau/r \sim 1/B \sim r^3$. 

A key result here is that, as expected, diffusion dominates near and just above the Kepler radius, while drift dominates at large radii, with the transition depending on the value of $\gamma$.
For fixed $\gamma$, the density is radially more extended and peaks at a larger radius for the constant diffusion/drift case (a=0) than the case ($a=3$) with radially increasing diffusion and drift.
The dominance of drift at large radii is particularly well illustrated by the log-log plots, for which the radial variation approaches a simple power law 
$1/r^{a+3} \sim 1/r^{b+2}$ at large radii, as shown by the good match in slope with the dashed, pure-power-law lines.

\begin{figure*}
\begin{center}
\includegraphics[scale=.45]{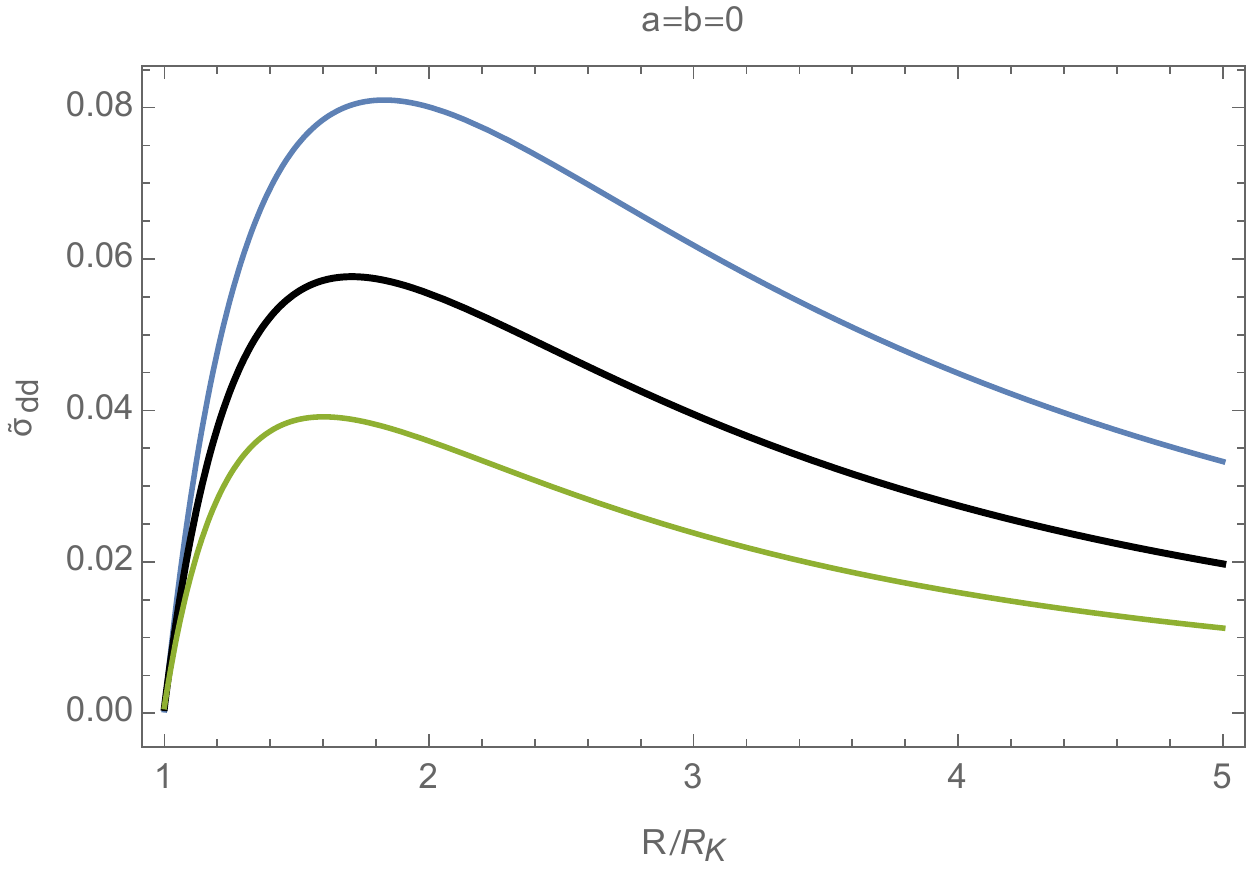}
~
\includegraphics[scale=.45]{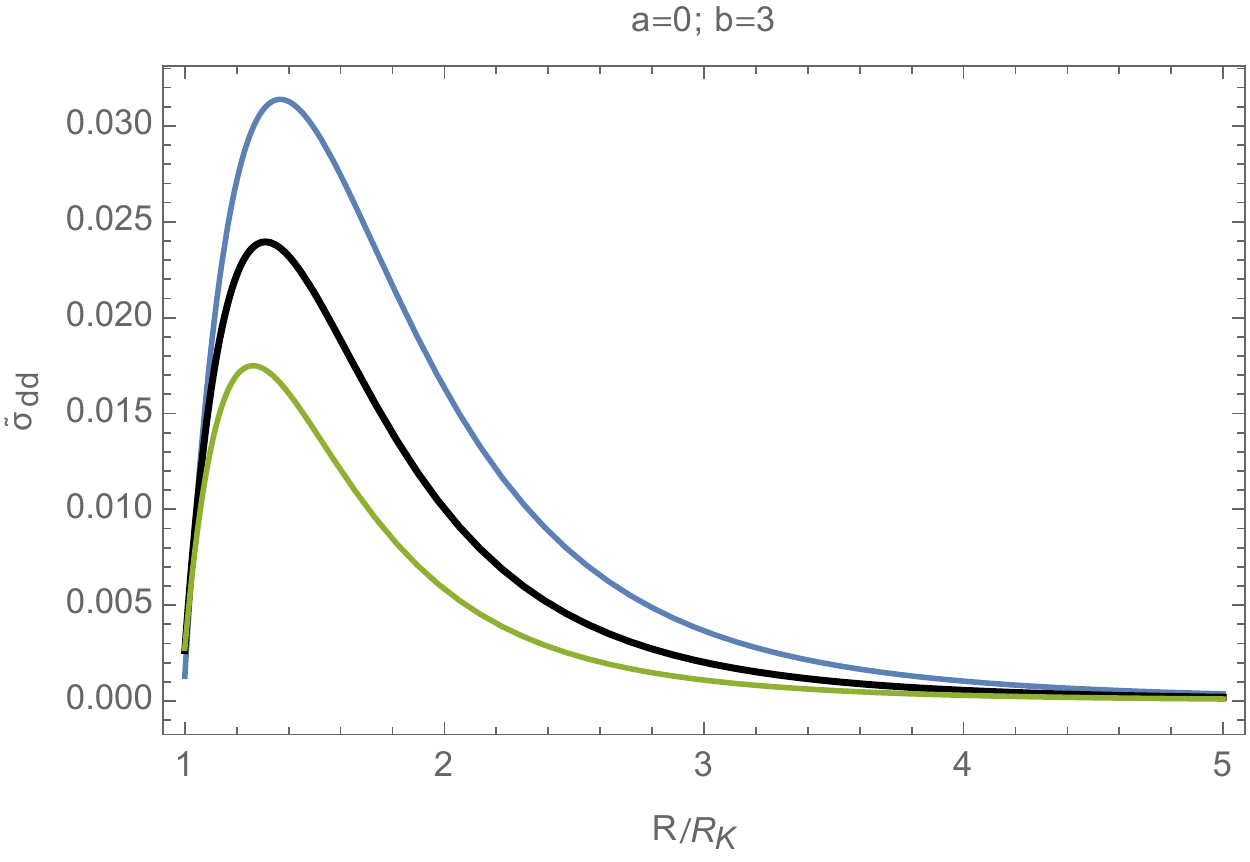}
~
\includegraphics[scale=.45]{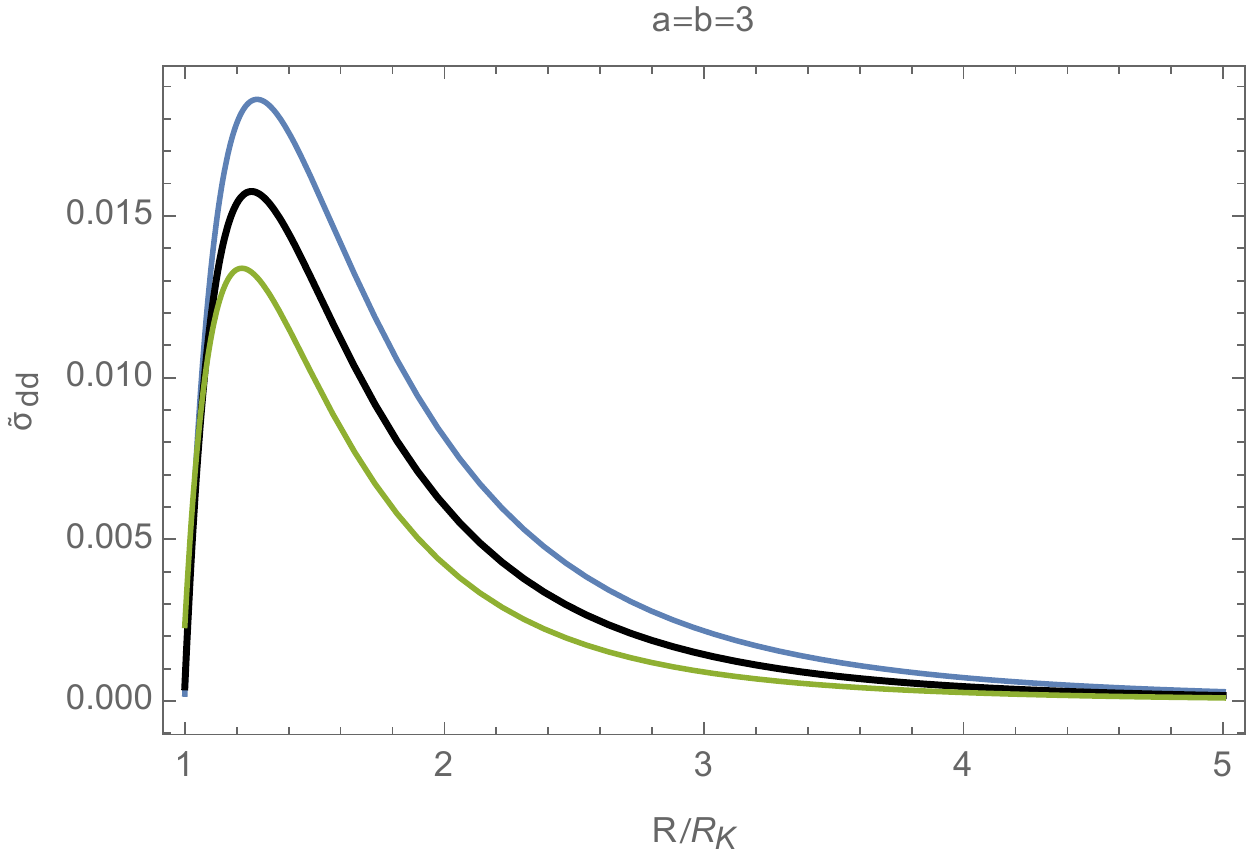}

\includegraphics[scale=.45]{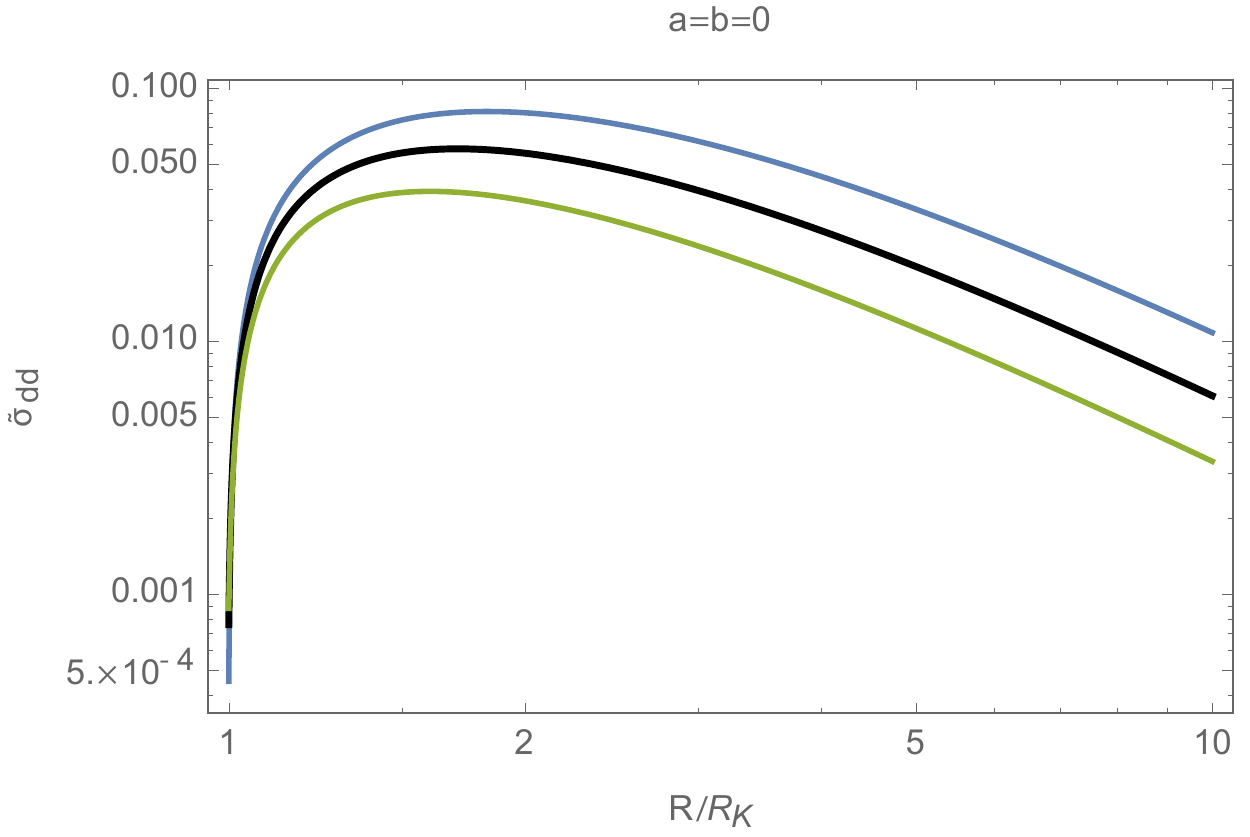}
~
\includegraphics[scale=.45]{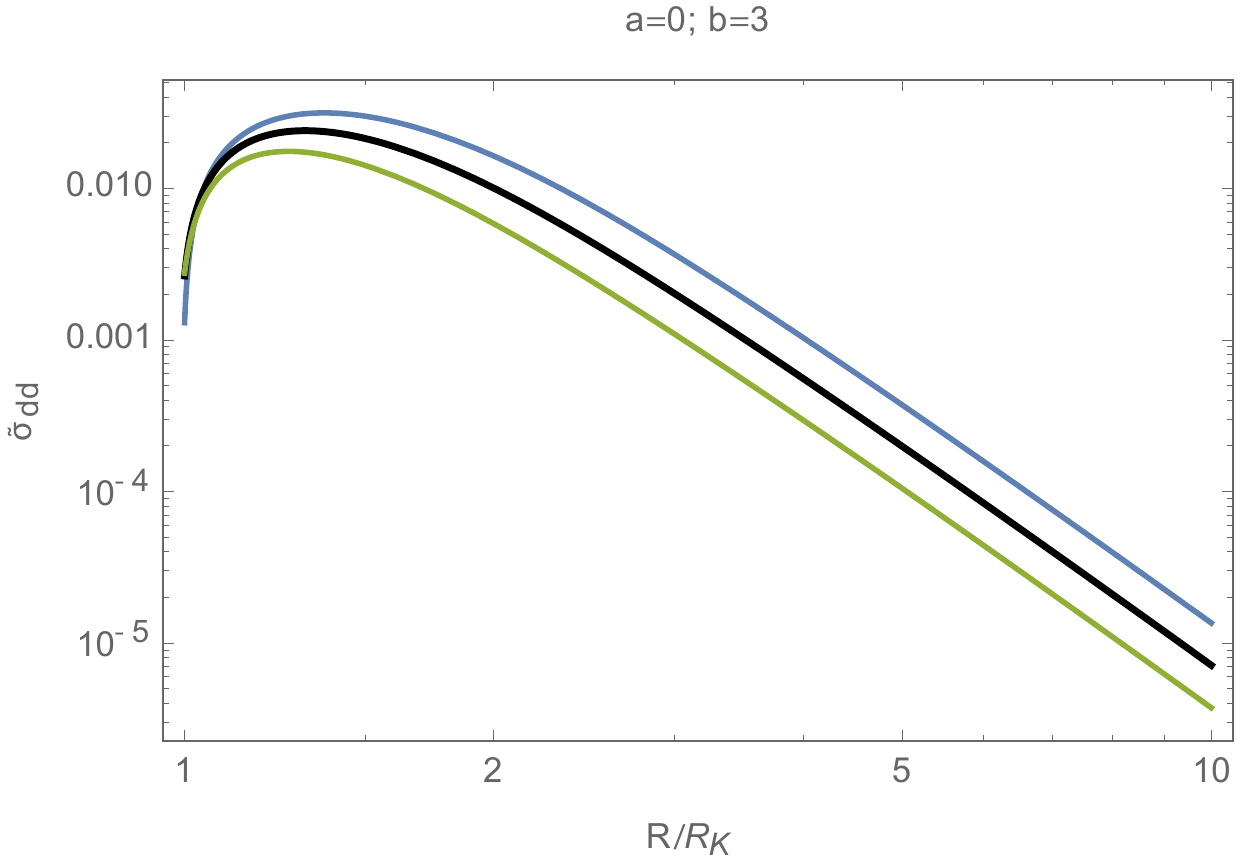}
~
\includegraphics[scale=.45]{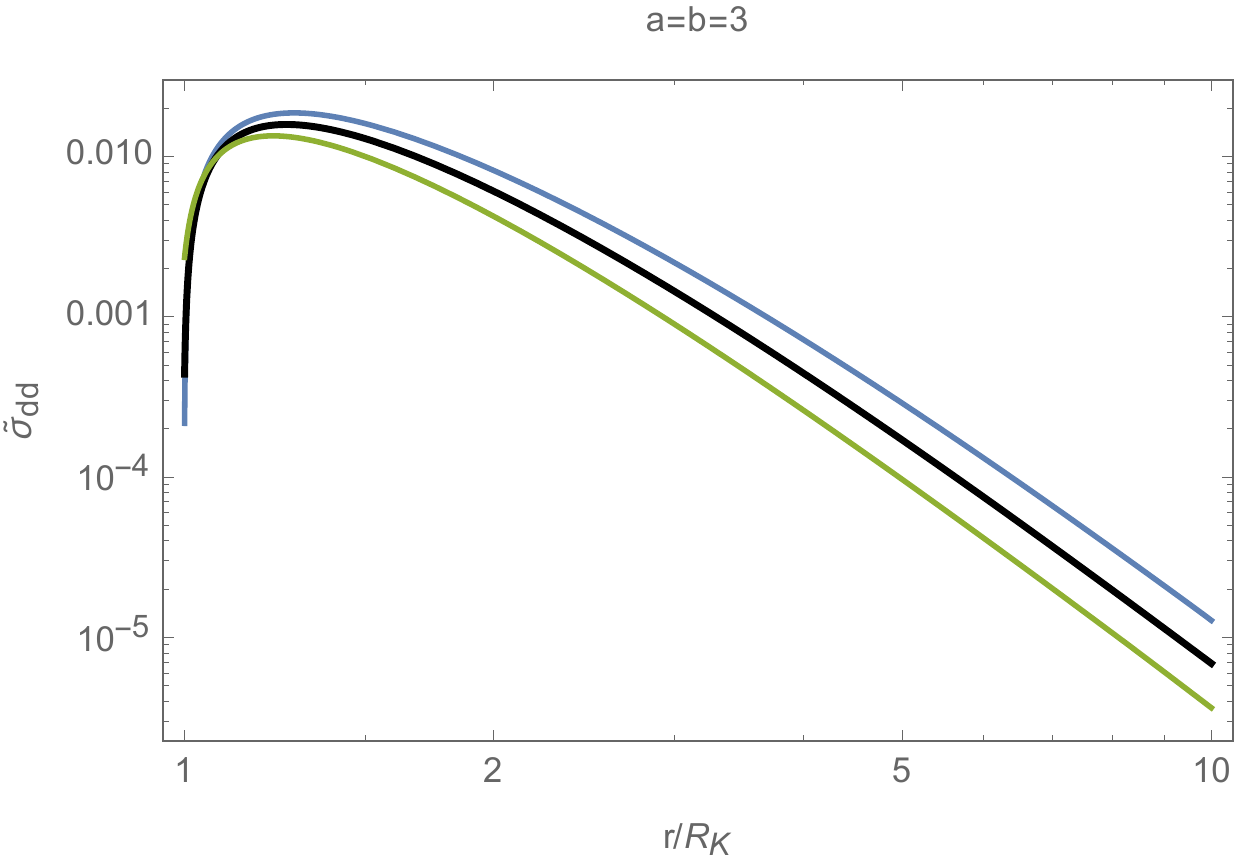}

\caption{As in figure \ref{fig:dd-solns}, but now for numerical solutions with labeled values of $a$ and $b$, and for $\gamma=1/2$, 1, and 2 (blue, black, green curves).
}
\label{fig:dd-nsolns}
\end{center}
\end{figure*}

Similar scalings can be derived numerically for general values of $b$. Figure \ref{fig:dd-nsolns} plots the radial variation of the scaled density for cases
$a=b=0$ (left), $a=0$ and $b=3$ (middle), and $a=b=3$ (right), with the top (bottom) row again showing linear (log) scaling.
The thick black line in each panel correspond to the $\gamma = 1$, while the blue (green) curves are for $\gamma=1/2$ ($\gamma =2$).
Just as in the analytic case with $b=a+1$, the models again all have a rounded variation near and above the Kepler radius, but approach a power-law variation $\tsigDd \sim 1/r^{b+2}$ in the drift-dominated regime at large radii.

\begin{figure*}
\begin{center}
\includegraphics[scale=.45]{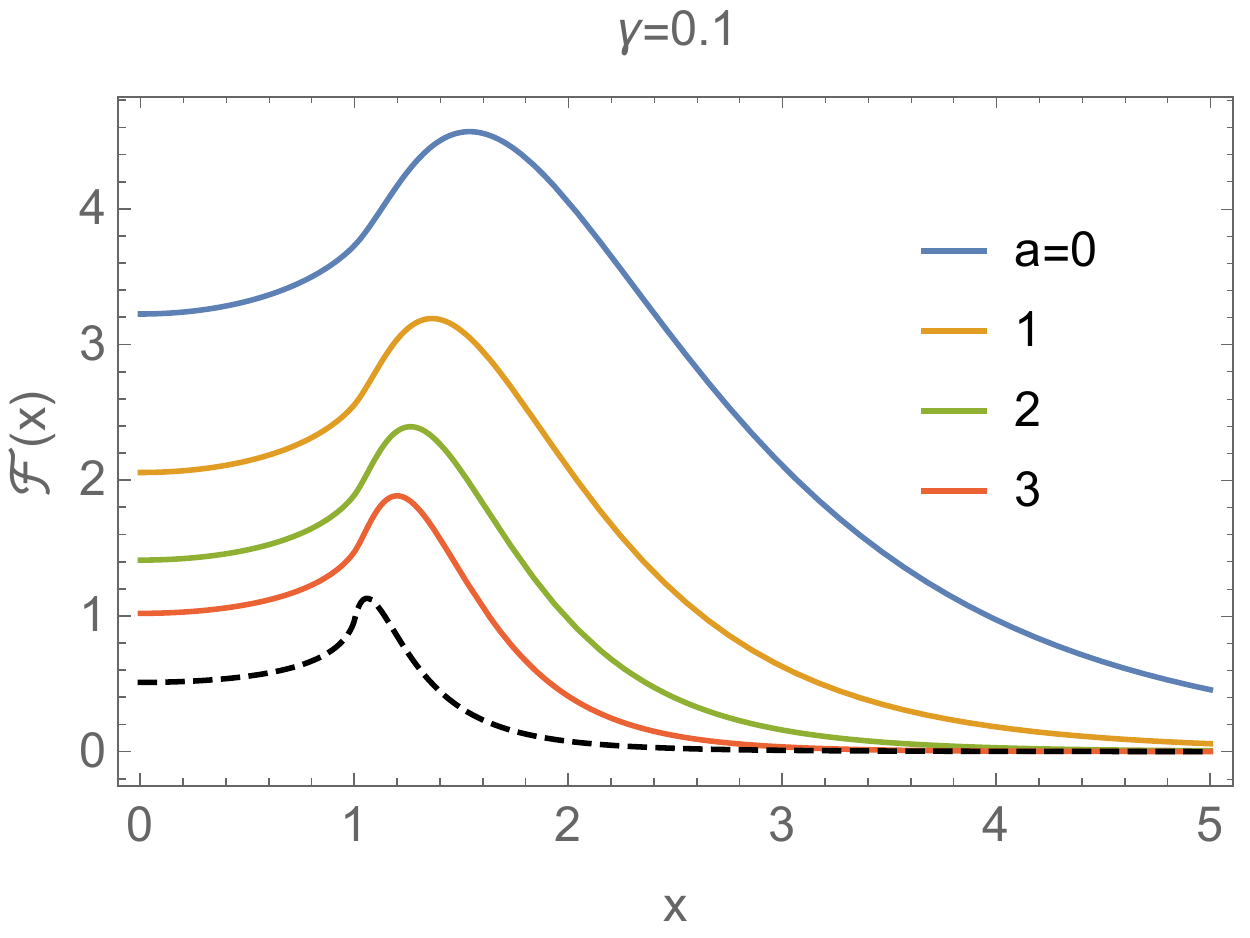}
~~
\includegraphics[scale=.45]{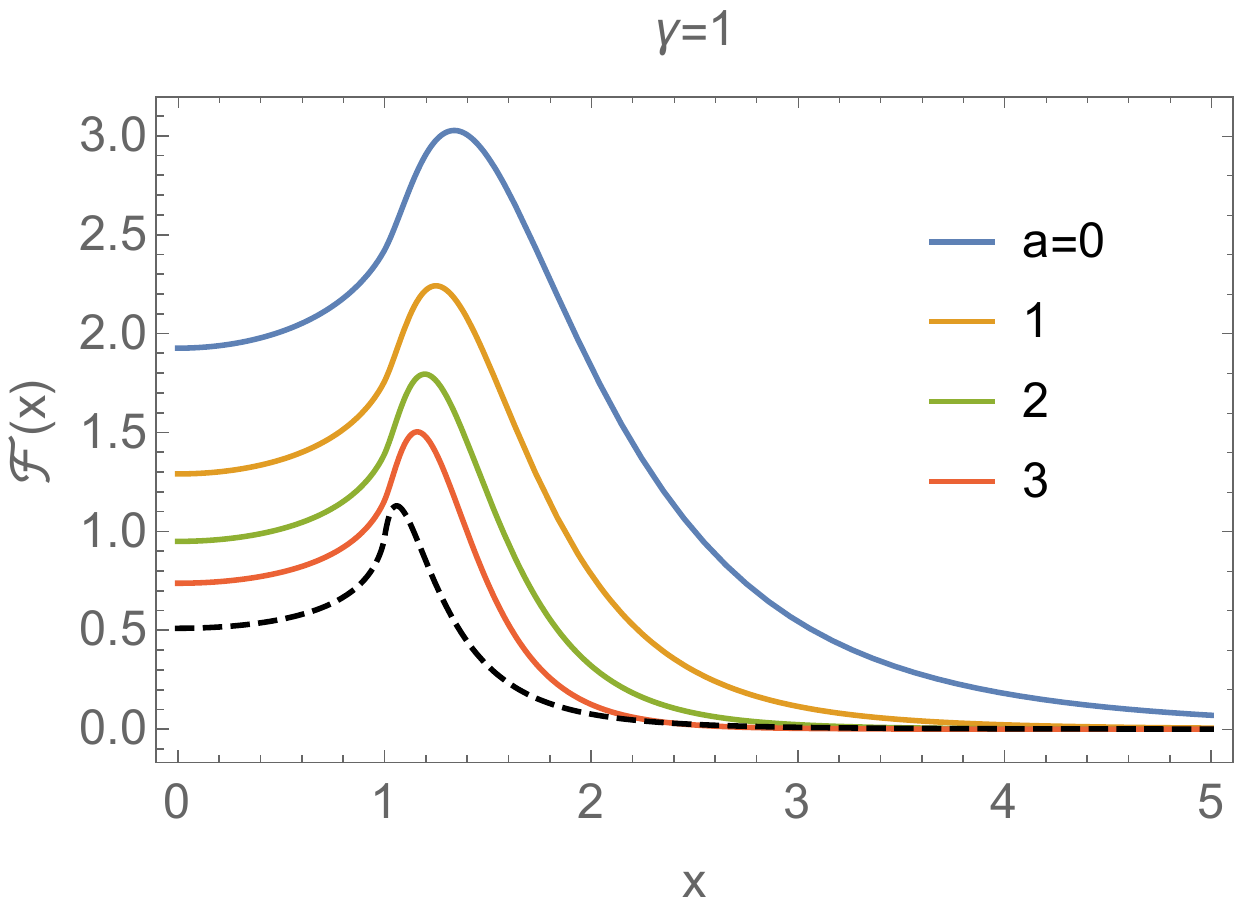}
~~
\includegraphics[scale=.45]{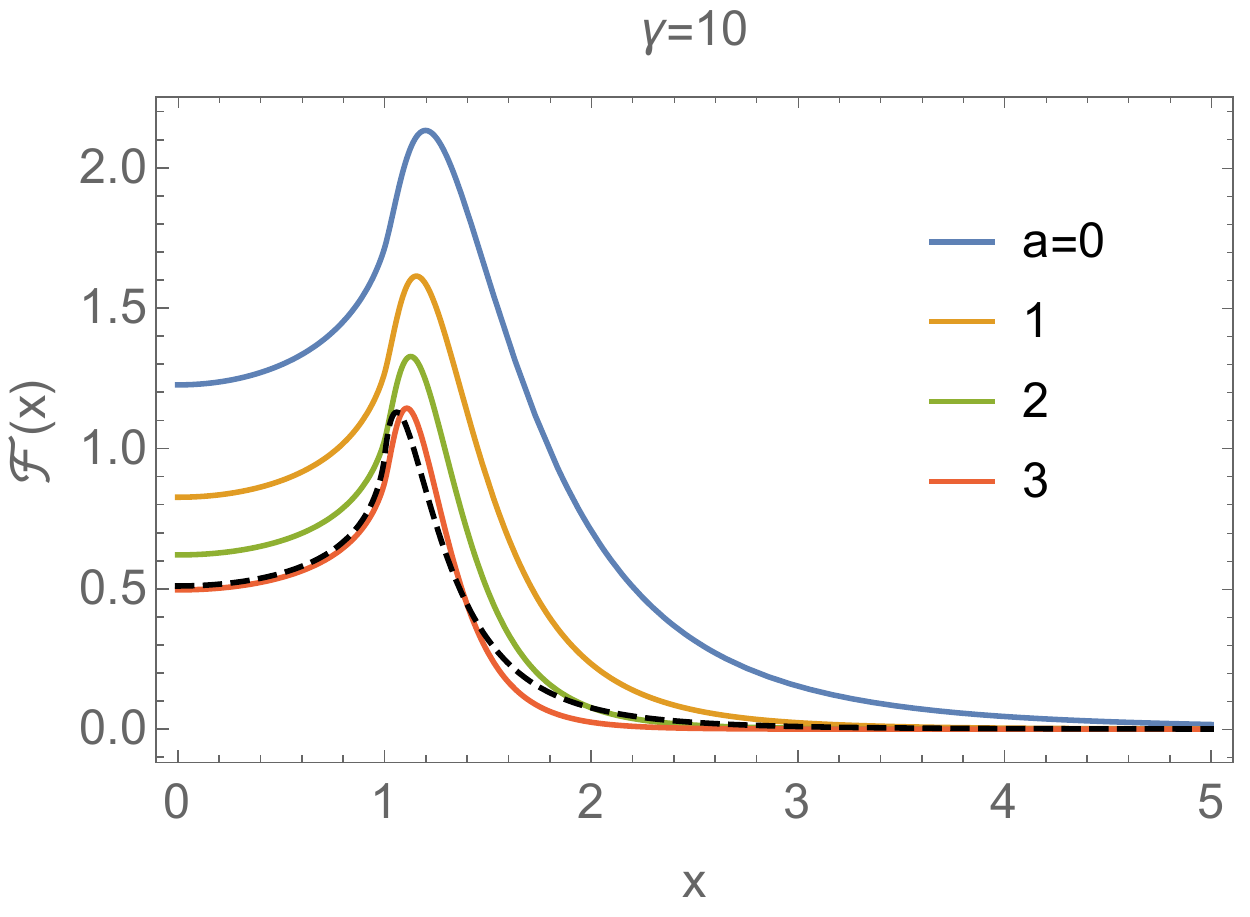}
\caption{
Line-emission flux profiles $\mathcal{F}(x)$ vs. $\sin i$ projected, Kepler-velocity-scaled wavelength $x$, computed from eqn.\ (\ref{eq:Fx}) for labeled values of drift-to-diffusion parameter $\gamma$, and diffusion radial power-indices $a=0$, 1, 2, and 3.
The emission peak is broader (narrower) for weaker (stronger) drift, as measured by the drift-to-diffusion parameter $\gamma$. 
At fixed $\gamma$, the peak location is at greater Doppler shift for models with radially constant diffusion ($a=0$) than in cases with radially increasing diffusion $a=3$.
But in all cases, the peak occurs only somewhat above the projected Doppler-shift ($x=1$) from co-rotation at the Kepler radius.
Note that the axisymmetry of the models imply a blue-red wind symmetry in the emission lines, so only one-half of the emission line profiles are plotted here.
The black dashed curves compare profiles for the RRM model, with disk density distributed according to the local feeding by the stellar wind, as given by eqn.\ \ref{eq:sigdef}.
The diffusion and drift models effects generally lead to profiles that have wider peaks than in the RRM model.
}
\label{fig:Fxg}
\end{center}
\end{figure*}

\begin{figure*}
\begin{center}
\includegraphics[scale=.65]{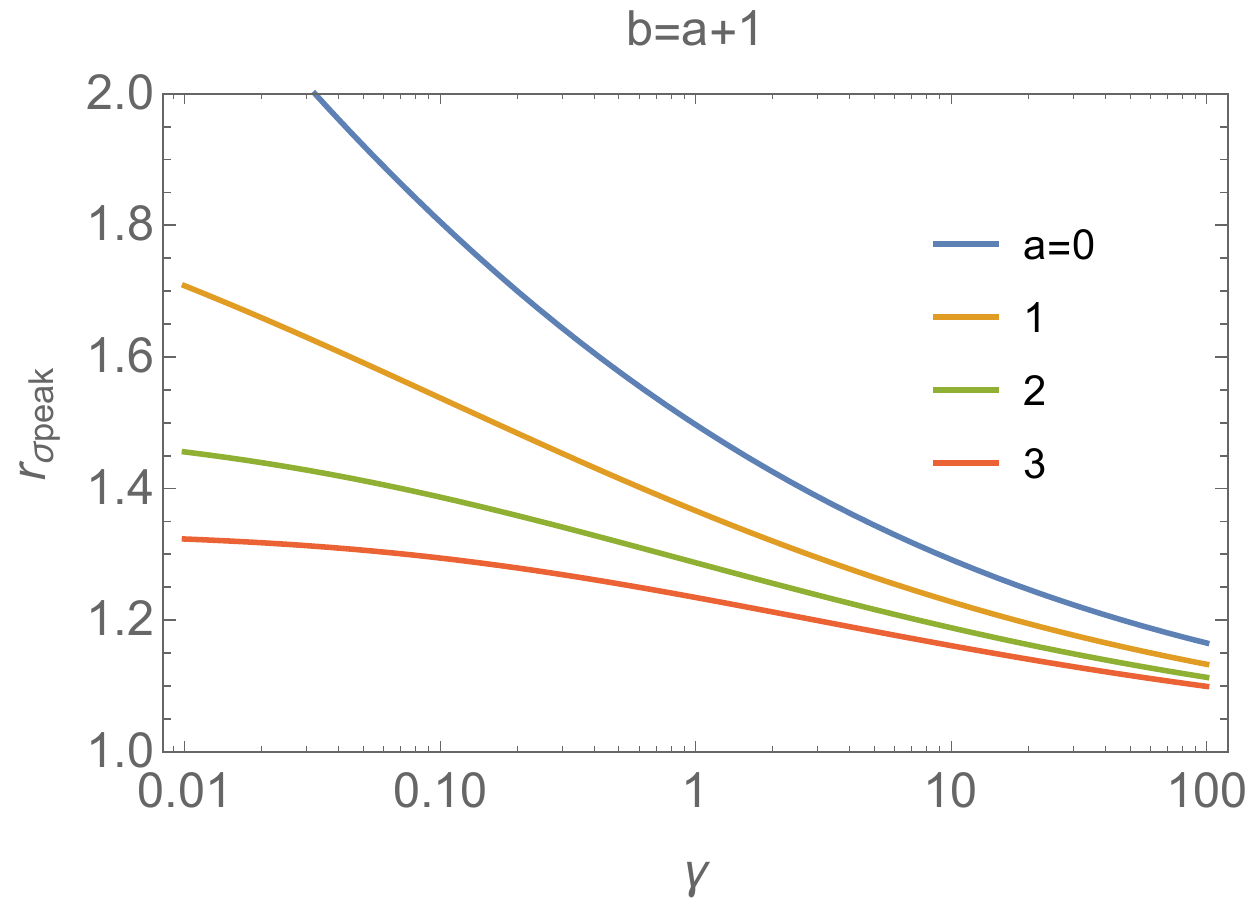}
\includegraphics[scale=.65]{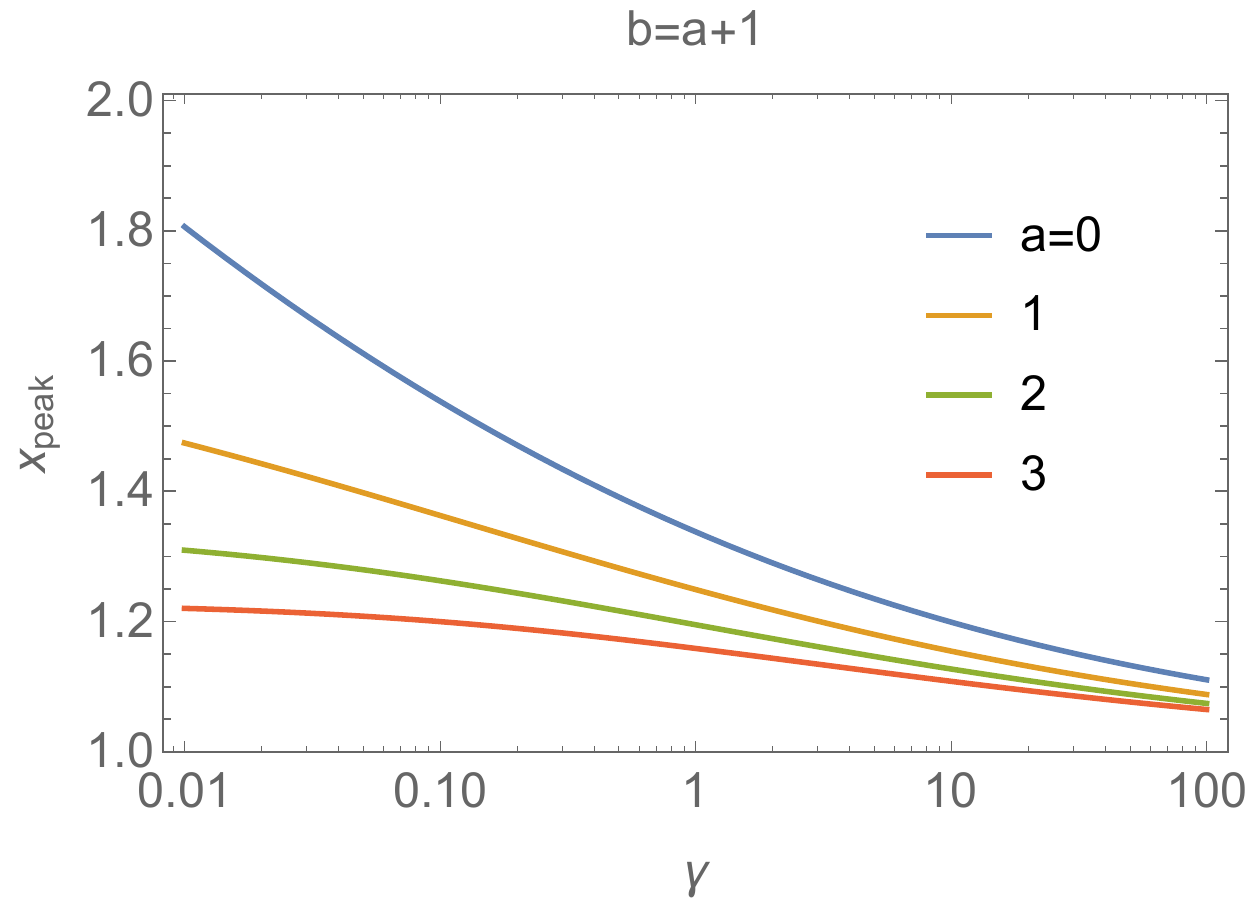}
\caption{Left: Kepler-radius-scaled radius of peak surface density vs.\ drift-to-diffusion parameter $\gamma$, for power indices  $a=0$, 1, 2, and 3.
Right: Kepler-velocity-scaled wavelength $x$ for peak flux vs. $\gamma$, for same power indices.
}
\label{fig:rxpeaks}
\end{center}
\end{figure*}

\begin{figure*}
\begin{center}
\includegraphics[scale=.65]{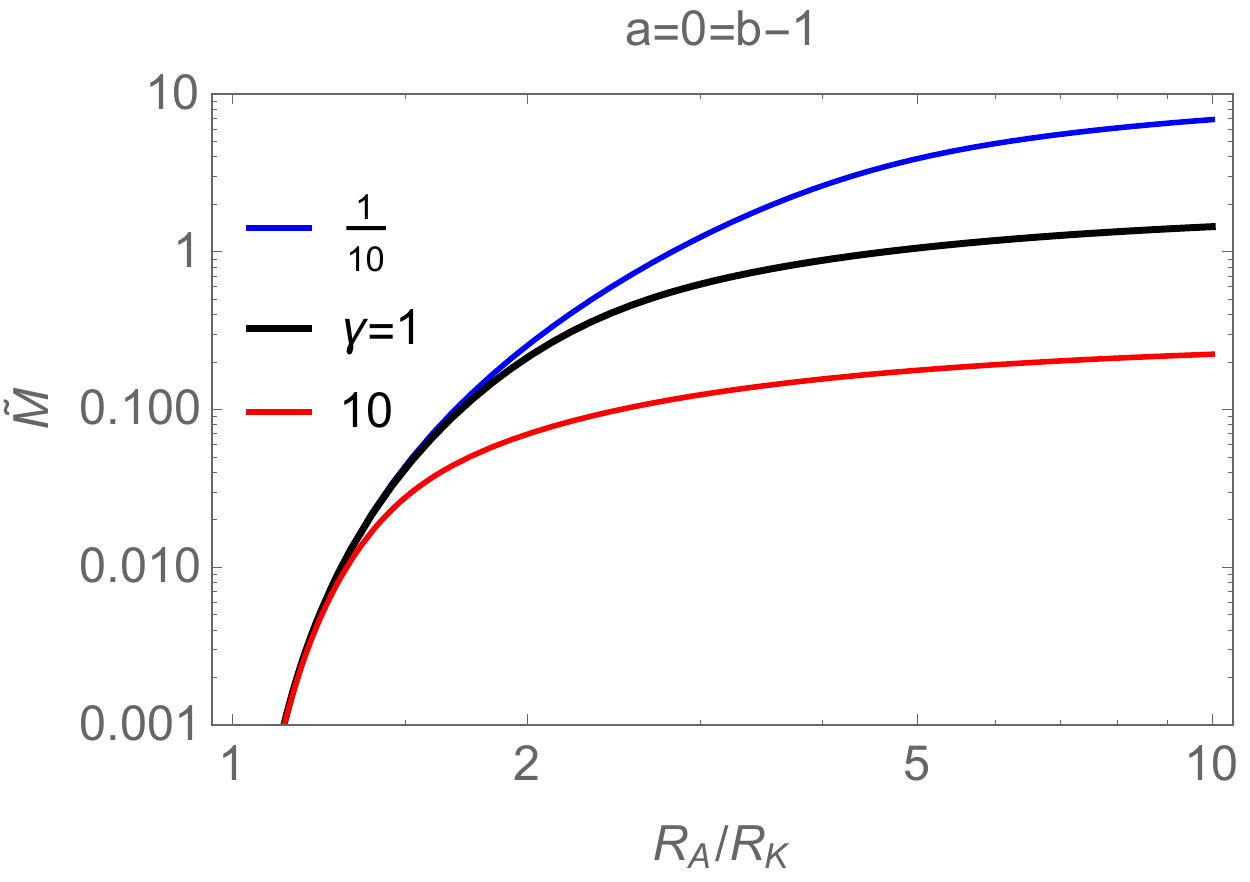}
~~
\includegraphics[scale=.65]{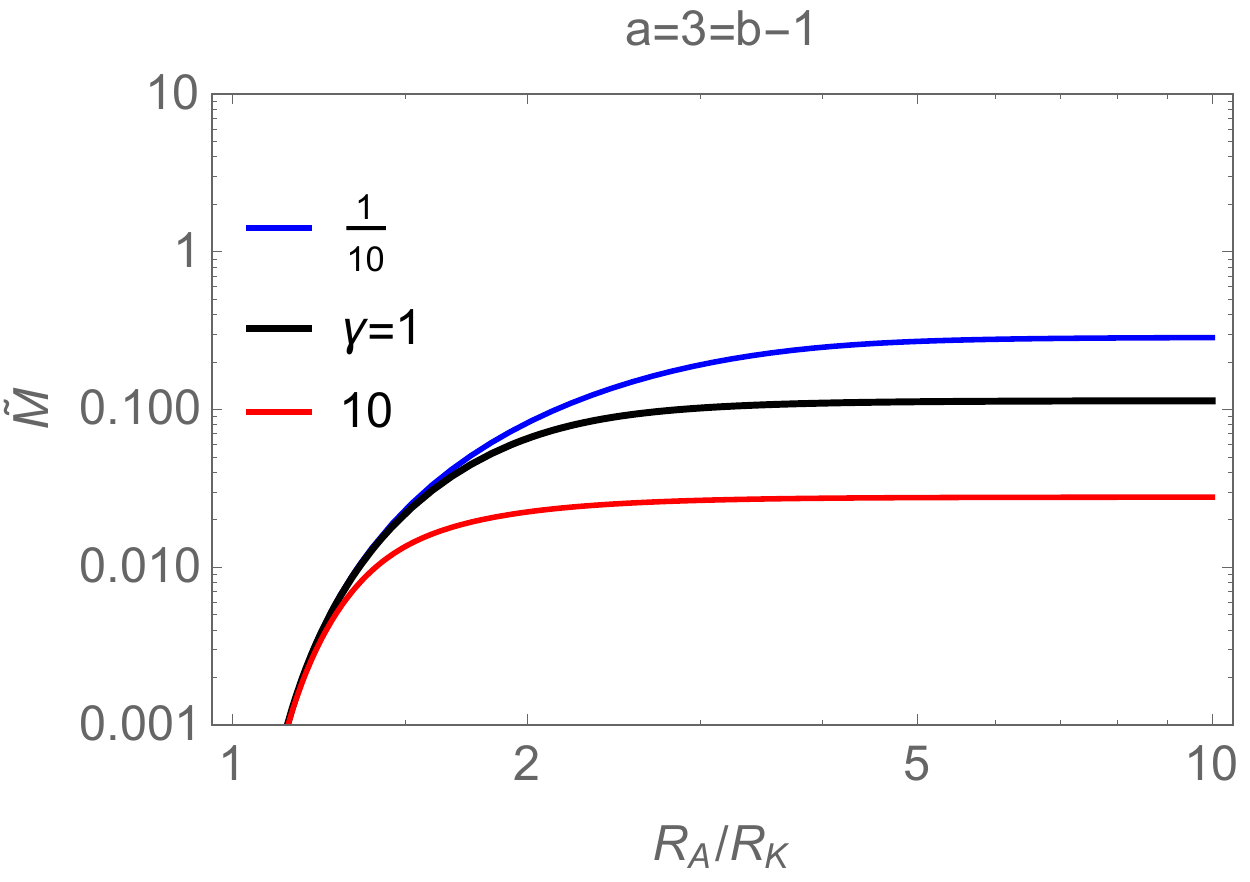}
\includegraphics[scale=.65]{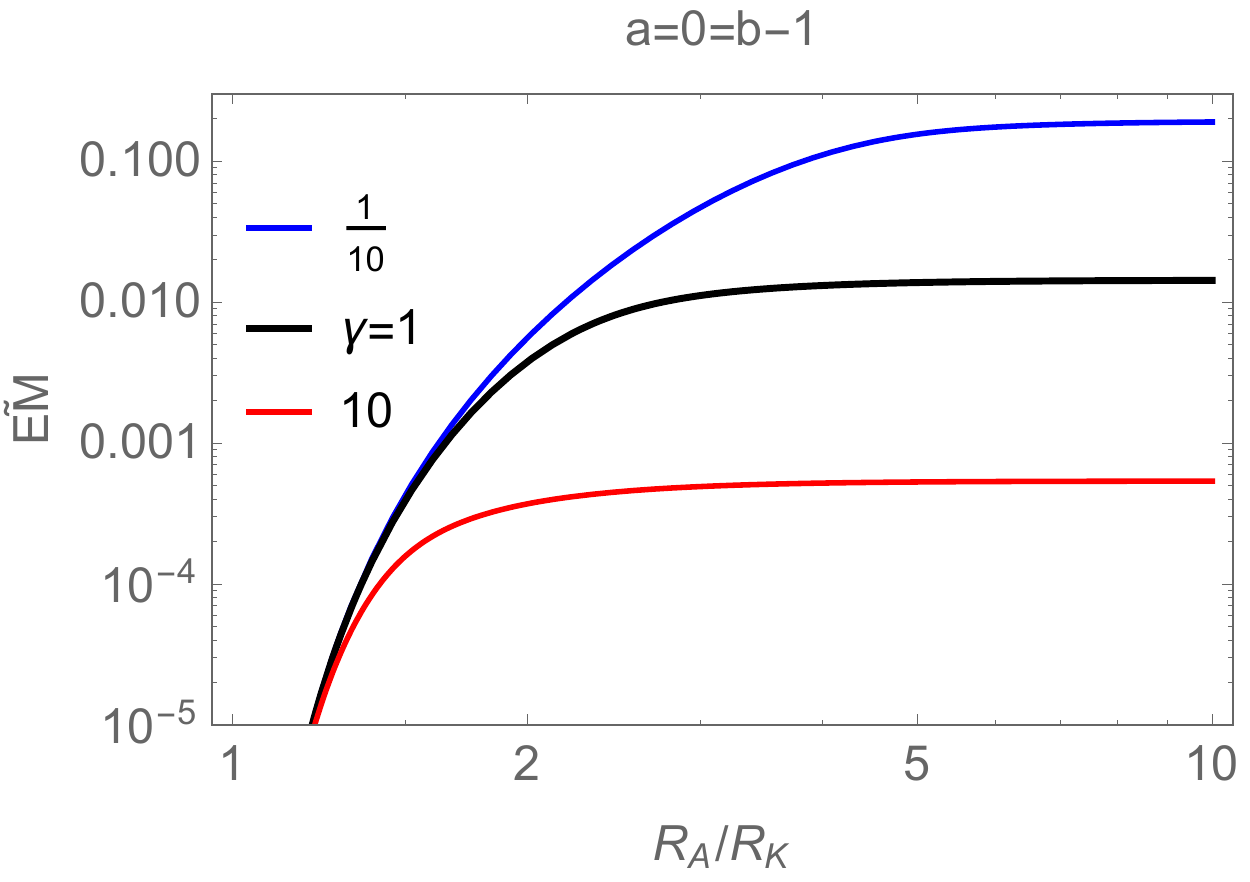}
~~
\includegraphics[scale=.65]{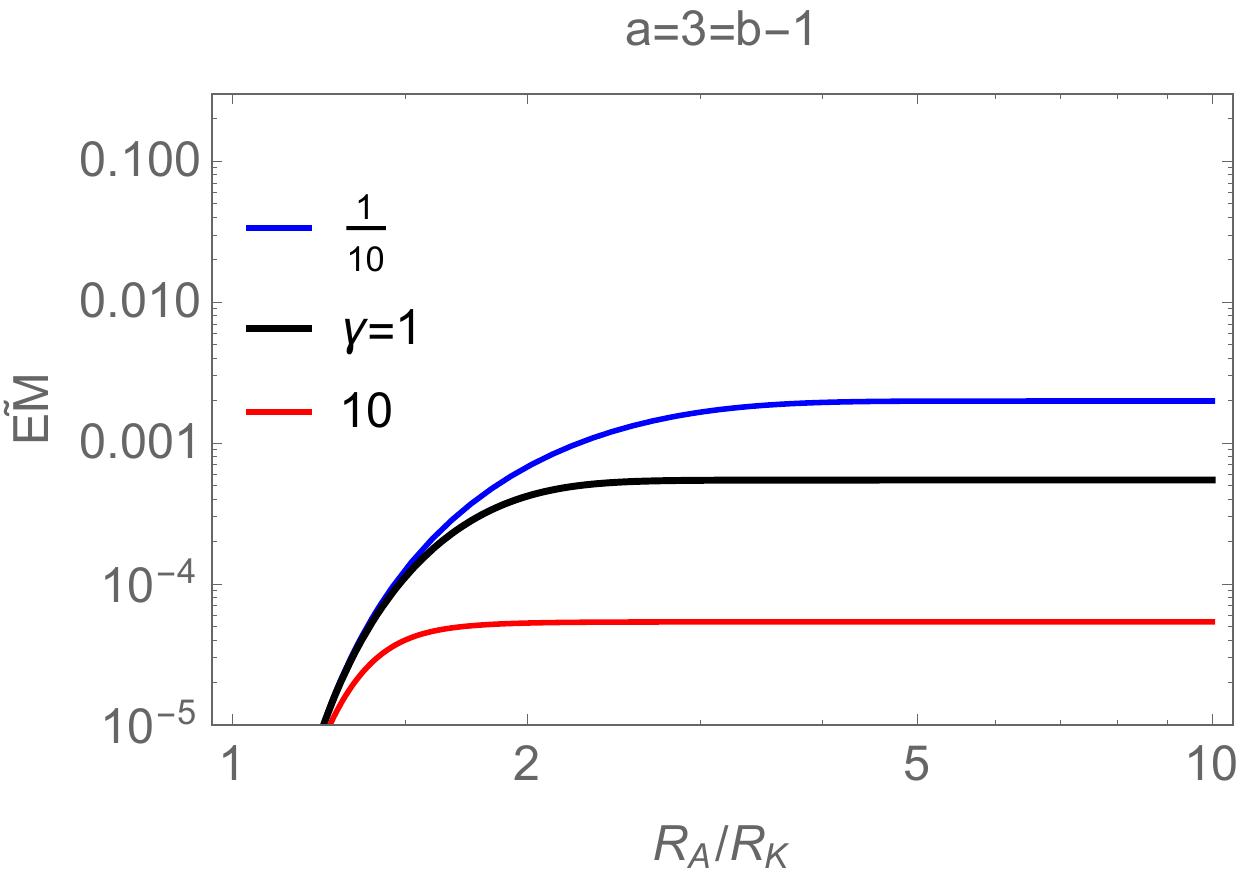}
\caption{Variation of the scaled mass ${\tilde M}$ (top) and scaled emission measure ${\tilde {EM}}$ (bottom) with increasing Alfv\'{e}n-to-Kepler radius ratio, $\rA=\RA/\RK$, for the same set of $\gamma$ values as in figure \ref{fig:dd-solns} (as labeled), and for the same two analytic cases $a=b-1=0$ (left) and $a=b-1=3$ (right).
}
\label{fig:mem-vs-ra}
\end{center}
\end{figure*}

\section{Observational diagnostics}

\subsection{Emission line profiles}

These diffusion+drift analytic solutions for the radial variation of the scaled surface density\footnote{For notational simplicity, we now drop the subscript ``Dd" and the explicit expression of the dependence on the parameters $\gamma$, $a$ and $b=a+1$} ${\tilde \sigma} (r)$ provide the basic information needed for a simple, optically thin model of emission line profiles.
Our approach here is basically the same as applied in TO05 eqn.\ (40). 
Namely, we assume the emission scales with the square of the local volume density $\rho (r) \sim \sigma(r)/h(r)$.

To proceed, consider  an observational coordinate system with the $z$ axis parallel to the observer's line of sight, the $x$ axis as a projected east--west impact parameter (i.e., perpendicular to the plane containing the stellar rotation axis), and the $y$ axis along the projected north-south direction.
For a rigidly rotating system, the Doppler-shifted wavelength from the line center varies only in the $x$ direction.
If we write $x$ and $z$ in units of the Kepler radius, then $r=\sqrt{x^2+z^2}$, with $x$ now also characterizing the wavelength Doppler shift in units of the projected velocity at the Kepler radius, $\vK \sin i = \Omega \RK \sin i $,  where $i$ is the usual inclination angle of the observer's line-of-sight to the rotation axis. 
The Doppler-shifted wavelength is thus given by $\lambda = \lambda_o (1 + x \Omega \RK \sin i /c)$. 
Within this scaling, the emission line profiles computed below (see figure \ref{fig:Fxg}) become independent of the inclination when plotted versus this wavelength $x$, which is scaled in term of the Doppler shift associated with the projected rotation speed at the Kepler radius, $\Omega \RK \sin i$.

Upon integration of the intensity over the $y$ coordinate (over which the line-of-sight Doppler shift is constant),
the observed flux emission line profile vs. $x$ is given (within a normalization constant) by\footnote{For $x< x_\ast \equiv R_\ast/\RK \sin i$, occultation from the star should truncate the integration for $z<0$; but to keep the results independent of the choice of $\RK/R_\ast$, we ignore this effect in the plots below.} 
further integration along the observer line-of-sight direction $z$,
 \beq
\mathcal{F} (x)  = 
 \int_{-\infty}^{\infty} \frac{\sigma^2(\sqrt{x^2+z^2})}{h(\sqrt{x^2+z^2})}  \, dz
 \, ,
 \label{eq:Fx}
 \eeq
 wherein the $y$-integration has reduced the net dependence on disk scale-height $h$ from inverse-squared to inverse-linear.
 
Figure \ref{fig:Fxg} plots $\mathcal{F}(x)$ vs. $x$ for $\gamma=0.1$, 1, and 10 (left, middle, right panels), and for selected radial power-indices $a=0$, 1, 2, and 3 
(blue, orange, green, red curves), with $b=a+1$.
The absolute normalization here is arbitrary; the plots were made by normalizing the surface density to its peak value for each case.

The emission peaks  are related to the radius of peak surface density.
The left panel of figure \ref{fig:rxpeaks} shows that this peak-density radius  ranges from $1.2 R_\ast$ to $2 R_\ast$, with higher values for lower power index ($a=0$) and lower 
diffusion-to-drift ratio parameter $\gamma$.
The right panel shows that the associated Doppler-shifted wavelengths $x$ for the peak emission flux have a similar variation, but over a somewhat smaller range.

Overall, the results show that the line profiles are broader, with a larger peak wavelength, for lower $a$ and smaller $\gamma$.

\begin{figure*}
\begin{center}
\includegraphics[scale=.65]{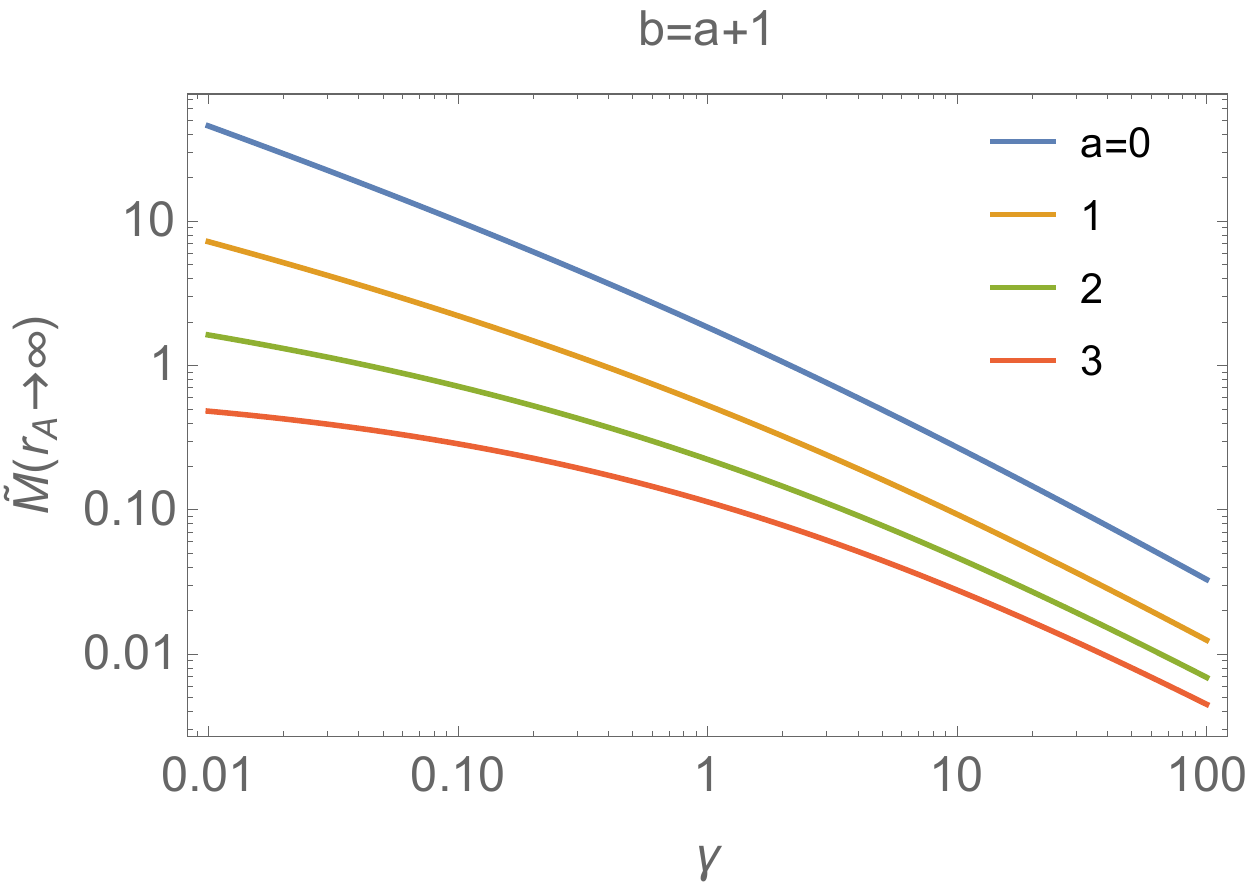}
\includegraphics[scale=.65]{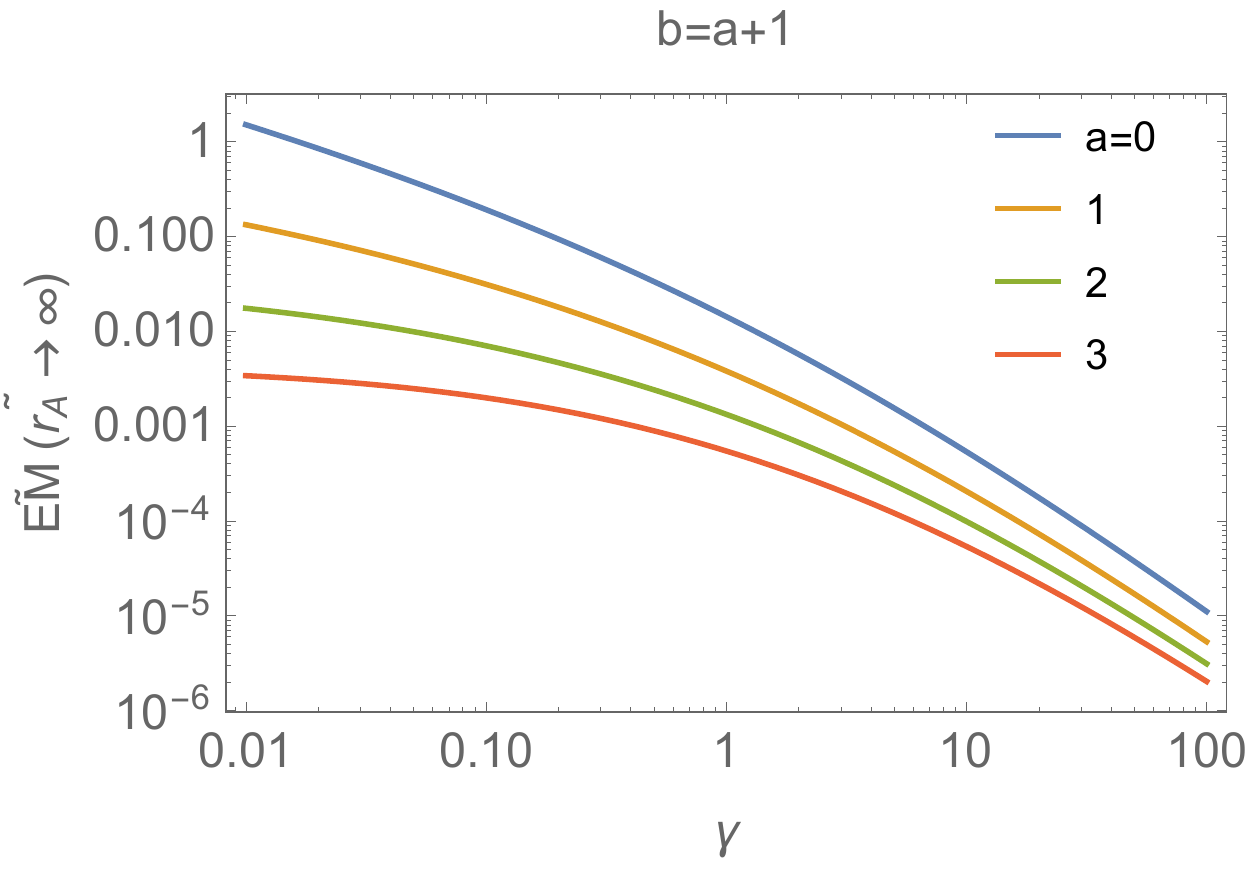}
\caption{Left: Scaled total mass in large-$\rA$ limit, plotted vs.\ drift-to-diffusion parameter $\gamma$, for power indices  $a=0$, 1, 2, and 3.
Right: Scaled total emission measure in large-$\rA$ limit, plotted  vs. $\gamma$, for same power indices.
}
\label{fig:MEMinf}
\end{center}
\end{figure*}

\subsection{Total Mass and Emission Measure}

Let us next derive scalings for the total mass $M$ of the magnetosphere,  and the associated emission measure $EM$.

In terms of the scaled disk density, the total mass between the Kepler and Alfv\'{e}n radii is given by the integral over surface area,
\beq
{\tilde M}
(\rA) = 2 \pi \int_1^{\rA} {\tilde \sigma} (r, \rA) \, r \, dr
\, .
\label{eq:ma}
\eeq
Likewise, the scaled {\em emission measure} between these radii is computed from
\beq
{\tilde {EM}}
(\rA) = 
2 \pi
\int_1^{\rA}  
{{\tilde \sigma}^2 (r,\rA)}{\sqrt{3-2/r^3}} 
~ r \, dr
\, ,
\label{eq:ema}
\eeq
where 
the square root factor accounts for slight radial variation in disk thickness, as given by eqn.\ (\ref{eq:hr}).

For the above analytic diffusion+drift models, figures \ref{fig:mem-vs-ra} plot the variation of this scaled mass (top) and scaled emission measure (bottom) with increasing Alfv\'{e}n-to-Kepler ratio, $\rA=\RA/\RK$, for the same set of $\gamma$ values as in figure \ref{fig:dd-solns}, and for the same two analytic cases $a=b-1=0$ (left) and $a=b-1=3$ (right).
Note that in all cases the curves increase from zero at $\RA=\RK$, but then flatten out for Alfv\'{e}n radii more than a few times the Kepler radius, albeit with a different asymptote, namely lower for larger $\gamma$ and/or larger $a$.  The first reflects the greater mass loss from stronger drift, while the latter stems from the more radially extended diffusion.

For a disk with given Kepler and Alfv\'{e}n radii, the associated dimensional form for the total mass $M(\RK,\RA)$ has the scaling
\beqa
\frac{M(\RK, \RA)} {{\tilde M} (\rA)}  &=& \sigDn \RK^2 
\nonumber
\\
&=& \frac{\sigdotK  \RK^4}{\DK}  
\nonumber
\\
&=& \frac{{\dot \sigma}_\ast R_\ast^4}{D_\ast} \left (\frac{\RK}{R_\ast} \right )^{1-a} 
\nonumber
\\
&=& \frac{{\dot M} \tau_{\mathrm D\ast}}{4 \pi} \left (\frac{\RK}{R_\ast} \right )^{1-a} 
\, ,
\label{eq:M}
\eeqa
where ${\dot \sigma}_\ast$ and $\tau_{\mathrm D\ast} \equiv R_\ast^2/D_\ast$ are respectively the disk feeding rate and  characteristic diffusion time at the scale of the stellar radius $R_\ast$.

In terms of the disk thickness at the Kepler radius, $\hK = \sqrt{2}\vs/\Omega$,
the dimensional emission measure $EM(\RK, \RA)$ has the scaling,
\beqa
\frac{EM(\RK, \RA)} {{\tilde {EM}} (\rA)}  &=& \frac{\sigDn^2}{\hK} \RK^2 
\nonumber
\\
&=& \frac{\sigdotK^2  \RK^6}{\hK \DK^2} 
\nonumber 
\\
&=& \frac{{\dot \sigma}_\ast^2 R_\ast^6}{\hK D_\ast^2} \left (\frac{\RK}{R_\ast} \right )^{-2a}
\nonumber 
\\
&=& \left ( \frac{{\dot M} \tau_{\mathrm D\ast}}{4 \pi} \right )^2 \frac{1}{\hK R_\ast^2}  \left (\frac{\RK}{R_\ast} \right )^{-2a} 
\, .
\label{eq:EM}
\eeqa

For any given set of diffusion and drift parameters, the full dimensional values for the mass and emission measure can be obtained from eqns.\ (\ref{eq:M}) and (\ref{eq:EM}), along with the associated scaled quantities. For the limit of large Alfv\'{e}n radii $\rA \rightarrow \infty$, figure \ref{fig:MEMinf} plots the variation of ${\tilde M}$ (left) and ${\tilde {EM}}$ (right) vs. $\gamma$, for the marked values of $a=0$, 1, 2, and 3.

For comparison, in {\em dynamical magnetospheres} (DM) with $\RA < \RK$, outflowing wind material trapped in loops that close within the Alfv\'{e}n radius falls back to the star on a dynamical timescale \citep{udDoula08}.
From the analytic dynamical magnetosphere (ADM) analysis of \citet{Owocki16},  the total mass $M_{\mathrm DM} (\RA)$ for a dynamical magnetosphere with 
 Alfv\'{e}n radius $\RA$ is given by
\beq
\frac{M_{\mathrm DM} (\RA)} {{\tilde M}_{\mathrm DM} (\RA)}  
=  \frac{{\dot M} R_\ast}{\Vesc} 
=  {\dot M} \tauff
\, ,
\label{eq:MDM}
\eeq
where $\Vesc = \sqrt{2GM_\ast/R_\ast}$ is the surface escape speed,  $\tauff=R_\ast/\Vesc$ is the associated free-fall time, 
and the scaled mass function ${\tilde M}_{\mathrm DM} (\RA)$ has a roughly linear scaling with $\RA-1$.

The appendix of TO05 also provides a discussion of how centrifugal breakout limits the maximum mass of a CM (see their eqn.\ A11),
\beqa
M_\infty &\approx& \frac{ \sqrt{\pi} }{6} {\dot M} \tauff \, \eta_\ast \left ( \frac{R_\ast}{\RK} \right )^2
\nonumber
\\
&\approx&  \frac{ \sqrt{\pi} }{12}  \frac{ B_\ast^2 R_\ast^2}{g_\ast}  \left ( \frac{R_\ast}{\RK} \right )^2
\, ,
\label{eq:TO05-A11}
\eeqa
where $\eta_\ast \equiv B_\ast^2 R_\ast^2/{\dot M} \Vesc^2$ is the disk magnetic confinement parameter defined in TO05 eqn.\ (A7).
Apart from order-unity factors, this confinement parameter, which can be of order $10^6$ or higher in many magnetic Bp stars with CM's, 
represents the key factor increase in this asymptotic mass of CM stars relative to the characteristic mass of DM stars, which eqn.\ (\ref{eq:MDM}) 
shows to be of order ${\dot M} \tauff$.
But because this parameter itself scales inversely with mass loss rate, the second equality shows that this asymptotic mass for centrifugal breakout is actually {\em independent} of ${\dot M}$.

\section{Diffusion and Drift in Turbulent Magnetic Fields}

Having derived these semi-analytical scalings for magnetospheric transport by 
diffusion and drift, let us next examine potential forms for the associated transport coefficients.
Out of the several different models proposed in other magnetospheric
contexts \citep[e.g.,][]{Fa66,HG84,Ac15},
we focus on the effects of turbulent cross-field diffusion and drift.
The surfaces of most OB stars have been inferred to exhibit stochastic
fluctuations over a range of spatial scales \citep{Ga96,Ae09}.
Although
 much  of the energy of these fluctuations may be
trapped in pulsational cavities, some fraction can ``leak out''
into the wind in the form of propagating waves.
If these waves induce transverse oscillations in the magnetospheric
field lines, then the plasma tied to an individual field line may
end up with a distribution of radial distances once it reaches the
magnetic equatorial plane.
This would be a manifestation of some combination of diffusion and drift.

The regions we have studied so far correspond to magnetically
dominated (high $\eta$ or low $\beta$) plasmas, so the corresponding
fluctuations are expected to fall into the {\em strong MHD turbulence}
regime.
This differs from traditional hydrodynamic turbulence \citep{Ko41,Ba53}
in several key ways.
Instead of being characterized by a single isotropic outer-scale
eddy size $\lambda$ and turnover timescale $\tau$ (with corresponding
eddy velocity $v \sim \lambda / \tau$), strong MHD turbulence exhibits
a strong wavenumber anisotropy, with a dominant cascade that takes
place in the two-dimensional plane perpendicular to the background field
\citep{St76,MT81,Sb83,HG84}.
The ``eddies'' appear to be composed mainly of Alfv\'{e}n-like wave
packets that propagate parallel to the field and interact with one
another nonlinearly only when they are counter-propagating
\citep{Ir63,Kr65}.

\citet{GS95} described the existence of a so-called ``critically
balanced'' state in which the perpendicular turnover time tends to
become the same order of magnitude as the Alfv\'{e}n wave period, with
\begin{equation}
  \tau_{\perp} \, = \, \frac{\lambda_{\perp}}{v_{\perp}} \, \approx \,
  \frac{\lambda_{\parallel}}{V_{\rm A}} \, = \, \tau_{\parallel}
\end{equation}
where $\lambda_{\parallel}$ and $\lambda_{\perp}$ are representative
correlation scales of the motions parallel and perpendicular to the
background field, $v_{\perp}$ is the root-mean-squared transverse eddy
velocity amplitude, and $V_{\rm A}$ is the Alfv\'{e}n speed of the
background field.
In such a system, a natural definition for the transverse coefficient of
``turbulent diffusion'' would be $D \sim \lambda_{\perp} v_{\perp}$,
and a natural drift timescale would be $\tau_{\perp}$ as defined above.

In the following subsections we investigate the scaling relations that
result from applying the above ideas to massive-star magnetospheres.

\subsection{Local Analysis}
\label{sec:local}

Transverse field-line wandering is usually described by a localized
spatial diffusion coefficient 
\begin{equation}
  \Delta \, = \, \frac{\langle \delta r^2 \rangle}{2 \, \delta s}
  \label{eq:dr2ods}
\end{equation}
which describes the transverse variance $\langle \delta r^2 \rangle$
of a bundle of field lines that originated at one common point $s$
and evolved turbulently to another point $s + \delta s$ along a direction
parallel to the background field.
There appear to be two limiting cases for $\Delta$ in the literature:
\begin{equation}
  \mbox{Quasilinear diffusion:} \,\,\,\,\,\,\,\,
  \Delta_q \, = \, \frac{\lambda_{\parallel} \, b^2}{B_0^2}
\end{equation}
\begin{equation}
  \mbox{Bohm diffusion:} \,\,\,\,\,\,\,\,
  \Delta_b \, = \, \frac{\lambda_{\perp} \, b}{B_0}
\end{equation}
where $B_0$ is the local background field strength and $b$ is
the root-mean-square amplitude of its transverse component
\citep[see, e.g.,][]{JP68,Ma95,Ru04,Sn15,Sn16}.
The quasilinear limit is often presumed to hold in the limit of
a low Kubo number:
$K = (\lambda_{\parallel} b)/(\lambda_{\perp} B_0) \lesssim 1$,
and the Bohm limit is believed to apply for $K \gtrsim 1$.
However, if the MHD turbulence exists in a state of near-equipartition
between transverse kinetic and magnetic perturbations, then
$b / B_0 \approx v_{\perp} / V_{\rm A}$, and the \citet{GS95}
condition of critical balance demands $K \approx 1$.
When that condition applies exactly, then $\Delta_q = \Delta_b$.

A true temporal diffusion coefficient can be estimated from the above
analysis by multiplying $\Delta$ by the propagation speed of fluctuations
along the field.
In regions well below the Alfv\'{e}n radius, the stellar wind speed
is negligible in comparison with $V_{\rm A}$, so we can write
\begin{equation}
  D \, = \, V_{\rm A} \Delta \, = \,
  \lambda_{\parallel} \, \frac{v_{\perp}^2}{V_{\rm A}}
  \label{eq:Dlocal}
\end{equation}
which is the same expression that would have resulted from applying the
simple scaling $D = \lambda_{\perp} v_{\perp}$ in the limit of critical
balance.
If the turbulence is carried by Alfv\'{e}n-like waves that originate
at the stellar surface and propagate along the field, the parallel
wavelength would be given by
\begin{equation}
  \lambda_{\parallel} \,\, \approx \,\, \frac{2\pi V_{\rm A}}{\omega}
  \label{eq:lambda_para}
\end{equation}
where the angular frequency $\omega$ is assumed to remain constant along
the field line.
For massive stars, the horizontal surface fluctuations that drive
magnetospheric $v_{\perp}$ motions are believed to originate in either
$p$-mode or $g$-mode oscillations.
Generally, $\omega \gtrsim 0.1 \omega_{\rm ac}$ for $p$-modes, where
$\omega_{\rm ac}$ is the photospheric acoustic cutoff frequency,
and $\omega \lesssim 0.01 \omega_{\rm ac}$ for $g$-modes
\citep{DeCat03,Cranmer09}.

Note that the transverse velocity amplitude $v_{\perp}$ appearing in
Equation (\ref{eq:Dlocal}) is not the value at the photosphere, but
instead should be evaluated locally at the apex of the magnetospheric
field line at radius $r$.
If the transverse fluctuations are similar enough to Alfv\'en waves,
wave-action conservation theory can be used to determine how $v_{\perp}$
changes in magnitude with increasing distance from the star
\citep[e.g.,][]{J77,IH82}.
Alfv\'en waves propagating into a magnetically dominated plasma
with stellar wind speed $u  \ll V_{\rm A}$ (i.e., for $r \ll R_{\rm A}$) 
exhibit $v_{\perp} = v_{\perp \ast} ( \rho_{\ast} / \rho)^{1/4}$,
with the subscript ``$\ast$'' indicating photospheric values.
Thus, the diffusion coefficient scales as
\begin{equation}
  D \, = \, \frac{2\pi v_{\perp}^2}{\omega} \, = \,
  \frac{2\pi v_{\perp \ast}^2}{\omega} \sqrt{\frac{\rho_{\ast}}{\rho}}
  \,\, .
\end{equation}
Because the field-line random walk is dominated by regions of the
magnetosphere exterior to the dense disk that forms at the midplane,
the density $\rho$ is that of the CAK stellar wind.
For regions far enough from the stellar surface that the stellar wind
has accelerated to a constant terminal speed,
the density scales as
$\rho(r) \propto B(r) \propto r^{-3}$, so the diffusion
coefficient $D \propto r^{3/2}$ (i.e., $a = 3/2$).
The normalization of $D$ at the Kepler radius depends on the driving
wave frequency and its photospheric velocity amplitude.

If the above description of turbulent diffusion is applicable to
massive stars, then the drift timescale should be comparable to the
critically balanced correlation times $\tau_{\parallel} \approx
\tau_{\perp}$.
The critical balance conjecture of \citet{GS95} implies that strongly
turbulent eddies remain coherent for roughly one characteristic
wave period, with $\tau \approx 2\pi / \omega$.
Given that this quantity is assumed to be constant as a function
of distance, the $b$ exponent from Section~2 would be zero.

\subsection{Global Analysis}
\label{sec:global}

The traditional field-line random walk theory tends to be applied
in regions for which the turbulence correlation scales $\lambda_{\perp}$
and $\lambda_{\parallel}$ are small in comparison with the macroscopic
size of the system.
This may not be true for massive-star magnetospheres.
For a B-type star like $\sigma$~Ori~E with fundamental parameters
equal to those given by \citet{To13} and a photospheric density
$\rho_{\ast} = 1.3 \times 10^{-9}$ g~cm$^{-3}$
\cite[computed by finding the location at which the Rosseland optical
depth is equal to one; see][]{Cr05},
the parallel correlation scale 
can be derived from Equation \ref{eq:lambda_para} as follows,
\begin{equation}
  \frac{\lambda_{\parallel}}{R_{\ast}} \, \approx \,
  0.35 \, \left( \frac{\omega_{\rm ac}}{\omega} \right)
  \left( \frac{R_{\ast}}{R} \right)^3 \,
  \sqrt{\frac{\rho_{\ast}}{\rho}}  \, \approx \,
  0.094 \sqrt{\frac{\rho_{\ast}}{\rho}}
\end{equation}
where the measured surface magnetic field strength ($B = 11$~kG) was
used to calculate the photospheric value of
$V_{\rm A} \approx 860$ km~s$^{-1}$, and the \citet{To13}
values of $M_{\ast}$, $R_{\ast}$, and $T_{\rm eff}$ were used to
compute $\omega_{\rm ac} \approx 0.006$ rad~s$^{-1}$.
The first expression above assumes that $B(R) \propto R^{-3}$, and
the second expression further assumes a representative value of the
frequency $\omega = 0.03 \omega_{\rm ac}$ and a radial distance
$r = 5 R_{\ast}$.
For a B-star mass loss rate of order $10^{-8}$ $M_{\odot}$~yr$^{-1}$,
the wind density at distances of 5--10 $R_{\ast}$ can be as low
as $10^{-17}$ to $10^{-15}$ g~cm$^{-3}$.
The resulting parallel turbulent length scale in that case would be
$\lambda_{\parallel} \approx 100$--1000 $R_{\ast}$.

In cases where $\lambda_{\parallel}$ is much larger than the radial
distance $R$, the diffusion may not have ample enough room to
develop to the degree predicted by the local analysis.
Thus, one possible solution could be to just replace
$\lambda_{\parallel}$ in Equation (\ref{eq:Dlocal})
by either $R$ itself or the field-line path length $s$.
For a dipole field, \citet{CS56} evaluated $s$ as a function of the
equatorial apex distance $R$.
Generally, $\omega \gtrsim 0.1 \omega_{\rm ac}$ for $p$-modes
and $\omega \lesssim 0.01 \omega_{\rm ac}$ for $g$-modes
\citep{DeCat03,Cranmer09}, where
$\omega_{\rm ac}$ is the acoustic cutoff frequency,
\begin{equation}
 \omega_{\rm ac} \, = \, \frac{\gamma g}{2 c_s} \,\, ,
\end{equation}
and the ratio of specific heats $\gamma = 5/3$, the gravitational
acceleration $g$, and the sound speed $c_s$ are all evaluated at the
photosphere.

The idea of replacing $\lambda_{\parallel}$ by either $R$ or $s$ 
is justifiable by computing the diffusion coefficient in a slightly
different way.
Equation (\ref{eq:dr2ods}) can be solved for the transverse spatial
variance of field lines in the magnetosphere by integrating along one
field line from the stellar surface to its apex; i.e.,
\begin{equation}
  \langle \delta r^2 \rangle \, = \, 2 \int ds \, \Delta(s)
  \, \approx \, 2.8 \int_{R_{\ast}}^{r} dr' \, \Delta(r')
\end{equation}
where the second expression is the result of applying the self-similar
coordinate transformation discussed above.  The resulting value of
$\langle \delta r^2 \rangle$ can then be divided by the critically balanced
timescale $\tau$ to obtain an estimate for $D$.
Note that the local expression for $\Delta$ is proportional to
$v_{\perp}^2 / V_{\rm A}$, and the density dependence
in each of those terms cancels out.
Thus, the only radial variation in $\Delta$ is an inverse dependence
on the field strength $B(R)$, and
\beqa
  D \, =\, \frac{\langle \delta R^2 \rangle}{\tau} \, 
  &\approx& \,
  2.8 \, v_{\perp \ast}^2 \sqrt{4\pi \rho_{\ast}}
  \int_{R_{\ast}}^{R} \frac{dR'}{B(R')} \, 
\nonumber
  \\
  &\approx& \,
  0.7 \, \frac{v_{\perp \ast}^2 R_{\ast}}{V_{{\rm A} \ast}}
  \left( \frac{R}{R_{\ast}} \right)^4
\eeqa
where $V_{{\rm A} \ast} = B_{\ast} / \sqrt{4\pi \rho_{\ast}}$.
The above expression is smaller by a factor of 2 than the result
of replacing $\lambda_{\parallel}$ by $s$ in the local definition of $D$.
Note that the radial dependence of this version of $D$ implies $a = 4$.
The normalization of $D$ at the Kepler radius depends on the velocity
amplitude and field strength at the photosphere.
The inverse dependence on $B_{\ast}$ makes intuitive sense,
because a stronger (stiffer) field ought to result in less
cross-field diffusion.

For the specific parameters of $\sigma$~Ori~E given above, it is
possible to estimate values for the diffusion and drift coefficients
as well as the time-steady magnetospheric mass.
Let us assume a photospheric transverse velocity
$v_{\perp\ast} \approx 10$ km~s$^{-1}$, which is close to the
measured macroturbulence velocity for this star \citep{Shultz16a}.
Combining that with other derived properties, such as
$R_{\rm K}/R_{\ast} = 2.5$ and $V_{{\rm A}\ast} = 860$ km~s$^{-1}$,
we obtain $D_{\rm K} \approx 10^7$ km$^2$~s$^{-1}$.
The numerator in the drift timescale
$\tau = \lambda_{\parallel}/V_{\rm A}$ should be replaced by $s$ if
considering the global limitations discussed above.
When evaluated at the Kepler radius, we find
$\tau_{\rm K} = s/V_{\rm A,K} \approx 120$ s.
These values give rise to $\gamma \approx 2$, which indicates roughly
comparable diffusion and drift at the Kepler radius.
Using $\tilde{M} \approx 0.1$, eqn.~(23) gives a mass of about
$6 \times 10^{-12} \, M_{\odot}$.
This is about an order of magnitude lower than the disk mass
estimated by \citet{To13}, and about a factor of 2000 lower
than the maximum breakout mass $M_{\infty}$.
Given the  approximate nature of this analysis, a mass
prediction of the right order is promising.

\subsection{Scaling Considerations}

Whether the turbulent diffusion is described using the local or
global approximations, the coefficient $D$ appears to depend on
the photospheric transverse velocity variance $v_{\perp \ast}^2$.
More intense wave activity means that a more vigorous field-line
random walk is ``seeded'' at the photosphere.
It may also be the case that $v_{\perp \ast}$ depends implicitly on
the star's dipole field strength $B_{\ast}$.
Magnetic fields may suppress the internal oscillations that are
sometimes observable as macroturbulence \citep{Su13}.
The near-surface field is also likely to have an impact on the degree of
mode conversion from internally trapped pulsations to freely propagating
MHD waves in the magnetosphere \citep[see also][]{MW68,Mu90,Je06}.

It may not be a coincidence that $D$ depends on some combination
of the wind density $\rho$ (as in Section \ref{sec:local}) and the
field strength $B_{\ast}$ (as in Section \ref{sec:global}).
This is somewhat similar to the interchange reconnection model
commonly invoked for planetary magnetosdisks.
\citet{Fa66}, \citet{SS81}, and others described the plasma content
of a given magnetic flux tube,
when intersecting the disk,
using a quantity
\begin{equation}
  \eta \, = \, \frac{\rho h}{B}
\end{equation}
that is nominally a field-line invariant similar to the magnetic moment.
However, \citet{Liu90} showed that interchange reconnection is
consistent with a slow cross-field diffusion, with
\begin{equation}
  \langle \delta R^2 \rangle \, \propto \,
  \frac{\partial \eta}{\partial R} \,\,\, ,
\end{equation}
in the disk
When the density $\rho$ in the definition of $\eta$ is assumed
to be that inside the cold magnetodisk, the problem becomes
fundamentally nonlinear (i.e., $D$ is a function of the solution
$\sigma$ to the diffusion equation).
However, in the field-line random walk picture, we interpreted $\rho$
as the density in the CAK stellar wind that feeds the disk.
If this is the case, it may motivate a diffusion coefficient that
scales as $D \sim \rho / B \sim {\dot M}/B$.

\section{Summary and Future Outlook}

This paper has explored diffusion+drift models for mass transport and leakage from stellar-wind-fed centrifugal magnetospheres around hot, luminous OB stars with moderate to rapid stellar rotation.
Within the assumption of field-aligned rotation, the semi-analytic results provide predictions for the relative mass distribution extending from the Kepler co-rotation radius $\RK$ to the wind Alfv\'{e}n radius $\RA$, and associated emission measure and line-emission profiles; it also shows how these scale with stellar and magnetic parameters, and with parameterizations for the diffusion and drift coefficients.
A key result is that, while wind mass-feeding is greatest near $\RK$,  diffusion limits the mass buildup in this region, in part  through inward leakage to the lower-density DM region below, but also from outward leakage to the drift region above.
The net outcome is that magnetospheric density peaks at radii distinctly above this Kepler radius.

An important area for future work will be to generalize this analysis to account for oblique rotators with a non-zero tilt between the rotation and magnetic axes.
With tilt, the mass accumulation surfaces become warped,  with most material accumulating in a plane roughly halfway between the rotational and magnetic equators.
Future work should examine how this can affect mass transport in a diffusion+drift model. 
One potential notion is that clouds that form at the intersection of the magnetic and rotational equators will have high density because they represent the points where the inner edge of the accumulation surface is closest to the star, and closest to the Kepler radius.
Further analysis is needed to work this out in detail, but it appears that at other azimuths, this inner edge is already above $\RK$. 
If so,  then it seems that such regions will be subject to centrifugal drift right from the inner edge, and so will have stronger mass leakage and thus lower density, 
with a radial drop off that is even steeper than the $r^{-3}$ scaling of TO05's fixed-time-accumulation models.

As such more general models are developed, further work will be needed to derive a broad array of observational diagnostics, including
polarization, photometric dips in continuum from cloud absorption, and the rotational modulation of emission-line profiles.

In a broader context, it would also be interesting to apply the analysis here to modeling the mass budget in planetary magnetosphere, e.g in the Jovian case in which the primary mass source from volcanic eruptions on the moon Io occurs well above the Kepler co-rotation radius
\citep{Bagenal97}. This would imply a dominance of outward drift effects over diffusion, with important consequences for the inferred mass distribution.

Finally, future work should also consider potential nonlinear feedback effects, for example with diffusion and/or drift coefficients that depend themselves on density; this could lead to a mass cap that is lower than predicted by direct centrifugal breakout, but depends mainly on field strength, and is independent of ${\dot M}$.

\section*{Acknowledgments}
SPO acknowledges support of visiting fellowships from the Laboratory for Atmospheric and Space Physics and from the Joint Institute for Laboratory Astrophysics at the University of Colorado, along with sabbatical leave support from the University of Delaware, and support from NASA grant NNX15AM96G.
SRCÕs work was supported by start-up funds from the Department of
Astrophysical and Planetary Sciences at the University of Colorado
Boulder.
We acknowledge helpful discussions with Fran Bagenal, Vero Petit, Matt Shultz,  Jon Sundqvist, Rich Townsend, Asif ud-Doula and Gregg Wade.

\end{document}